\documentclass[a4paper,11pt]{article}
\pdfoutput=1 % if your are submitting a pdflatex (i.e. if you have
\usepackage{jheppub} % for details on the use of the package, please
\usepackage{graphicx}% Include figure files
\usepackage{txfonts}% this package should be put before bm
\usepackage{bm}% bold math

\usepackage{amsfonts}
\usepackage{amsmath}
\usepackage{subfigure}
\usepackage{inputenc}
\usepackage{graphics}
\usepackage{dsfont}
\usepackage{mathtools}
\usepackage{autobreak}
\usepackage{diagbox}
\usepackage{multirow}
\usepackage{longtable}
\usepackage{etoolbox}%avoid the underfull \hboxof bib
\apptocmd{\sloppy}{\hbadness 10000\relax}{}{}

\usepackage{overpic}
\usepackage{indentfirst}
\usepackage{cancel} %for Feynman symbols
\usepackage{slashed}  %for Feynman symbols
\usepackage{cases}
\usepackage{color}
\usepackage{natbib}
\allowdisplaybreaks[4]
\newcommand{\GGb}{\langle \alpha_s GG\rangle}

\newcommand{\GGGb}{\langle g_s^3G^3\rangle}

\newmuskip\pFqskip
\mathchardef\pFcomma=\mathcode`, % keep a copy of the comma

\newcommand*\pFq[5]{%
  \begingroup
  \begingroup\lccode`~=`,
    \lowercase{\endgroup\def~}{\pFcomma\mkern\pFqskip}%
  \mathcode`,=\string"8000
  {}_{#1}F_{#2}\biggl(\left.\genfrac..{0pt}{}{#3}{#4} \right| \,#5\biggr)%
  \endgroup
}

\newmuskip\MeijerGskip
\mathchardef\pFcomma=\mathcode`, % keep a copy of the comma

\newcommand*\MeijerG[5]{%
  \begingroup
  \begingroup\lccode`~=`,
    \lowercase{\endgroup\def~}{\pFcomma\mkern\pFqskip}%
  \mathcode`,=\string"8000
  G^{#1}_{#2}\biggl( #5\left|\genfrac..{0pt}{}{#3}{#4} \right.\biggr)%
  \endgroup
}
\title{Revisit the heavy quarkonium double-gluon hybrid mesons with exotic quantum numbers}

\author[a]{Ding-Kun Lian,}
\author[a]{Qi-Nan Wang,}
\author[a]{Xu-Liang Chen,}
\author[a]{Peng-Fei Yang,}
\author[a,b]{Wei Chen}
\author[c]{and Hua-Xing Chen}

\affiliation[a]{School of Physics, Sun Yat-Sen University, Guangzhou 510275, China}
\affiliation[b]{Southern Center for Nuclear-Science Theory (SCNT), Institute of Modern Physics, 
Chinese Academy of Sciences, Huizhou 516000, Guangdong Province, China}
\affiliation[c]{School of Physics, Southeast University, Nanjing 210094, China}

\emailAdd{liandk@mail2.sysu.edu.cn}
\emailAdd{wangqn8@mail.sysu.edu.cn}
\emailAdd{chenxliang@mail2.sysu.edu.cn}
\emailAdd{yangpf5@mail2.sysu.edu.cn}
\emailAdd{chenwei29@mail.sysu.edu.cn}
\emailAdd{hxchen@seu.edu.cn}

\abstract{We revisit the masses of heavy quarkonium double-gluon hybrid mesons with exotic quantum numbers $J^{PC}=1^{-+}$ and $2^{+-}$ in the framework of the QCD sum rules. Considering the double-gluon hybrid meson operators in the octet-octet color structure, we have constructed two independent interpolating currents with $J^{PC}=1^{-+}$ and five independent currents with $J^{PC}=2^{+-}$. For the interpolating currents with antisymmetric glueball operator, there exist non-local divergences in one kind of additional Feynman diagrams of the tri-gluon condensate, which will give important contributions to the sum rule stabilities and mass predictions. We use the diagrammatic renormalization to cancel out such divergences. At the leading order of $\alpha_s$, the two-point correlation functions and spectral densities can be expressed in the analytic form of the generalized hypergeometric functions and Meijer's G-functions. 
After performing the numerical analysis, we predict the masses of the $1^{-+}$ and $2^{+-}$ charmonium double-gluon hybrid mesons to be around $6.1-7.2$ GeV and $6.3-6.4$ GeV, respectively. For the bottomonium systems, their masses are predicted to be $13.7-14.3$ GeV and $12.6-13.3$ GeV for the $1^{-+}$ and $2^{+-}$ channels, respectively. Besides, it is possible to hunt for these charmonium hybrids in the radiative decays of bottomonium mesons in BelleII experiment. Further investigations on these hybrid states in various theoretical and phenomenological methods are also anticipated in the future.}

\arxivnumber{2403.18696}

\begin{document}
   
\maketitle
\flushbottom
%================================================================================
\section{Introduction}
%================================================================================
\noindent Hadrons are $q\bar{q}$ mesons and $qqq$ baryons in the conventional quark model (CQM)~\cite{ParticleDataGroup:2022pth,Gell-Mann:1964ewy,Zweig:570209}. However, there exist exotic hadrons in quantum chromodynamics (QCD) beyond CQM, such as multiquark states, hybrid mesons and glueballs, etc. A typical hybrid meson is composed of a pair of quark-antiquark and one valence gluon ($\bar{q}Gq$). Due to the existence of gluonic degree of freedom, hybrid mesons can carry exotic quantum numbers ($J^{PC}=0^{--},even^{+-},odd^{-+}$), which are forbidden in the ordinary $q\bar{q}$ systems. To date, the most fascinating candidates of light hybrid mesons are $\pi_1(1400)$~\cite{IHEP-Brussels-LosAlamos-AnnecyLAPP:1988iqi}, $\pi_1(1600)$~\cite{E852:2001ikk} and $\pi_1(2015)$~\cite{E852:2004gpn} with the exotic quantum number $I^{G}J^{PC}=1^{-}1^{-+}$. Recently, an isoscalar $\eta_{1}(1855)$ was reported by BESIII with $I^{G}J^{PC}=0^{+}1^{-+}$ in the decay process $J/\psi\rightarrow \gamma\eta_{1}\rightarrow \gamma\eta\eta^{\prime}$~\cite{BESIII:2022riz,BESIII:2022iwi}, which has been  considered as a partner state of $\pi_1(1600)$ in a hybrid nonet~\cite{Chen:2022qpd,Chen:2023ukh,Chen:2022isv,Qiu:2022ktc,Shastry:2023ths,Shastry:2022mhk}. 

In the past several decades, there have been many theoretical studies on the $\bar{q}Gq$ hybrid mesons based on various methods, such as the MIT bag model~\cite{Barnes:1982tx,Chanowitz:1982qj}, flux-tube model~\cite{Close:1994hc,Barnes:1995hc,Page:1998gz}, lattice QCD~\cite{McNeile:1998cp,Lacock:1998be,Bernard:2003jd,Hedditch:2005zf,Dudek:2009qf,Dudek:2010wm,Dudek:2011tt,Dudek:2011bn} and QCD sum rules~\cite{Balitsky:1982ps,Govaerts:1983ka,Govaerts:1984hc,Govaerts:1985fx,Govaerts:1986pp,Zhu:1998ki,Jin:2002rw,Qiao:2010zh,Berg:2012gd,Chen:2013zia,Chen:2013pya,Chen:2013eha,Li:2021fwk,Palameta:2017ols,Palameta:2018yce,Ho:2018cat,Ho:2019org,Chen:2022qpd}. It is interesting that there exists a supermultiplet structure for the $\bar{q}Gq$ hybrid mesons, among which the states with $J^{PC}=(0,1,2)^{-+},\, 1^{--}$ form the lightest hybrid supermultiplet, while the states with $J^{PC}=(0,2)^{+-}$ constitute a heavier hybrid supermultiplet~\cite{HadronSpectrum:2012gic,Chen:2013zia,Meyer:2015eta,Dudek:2013yja,Dudek:2009qf,Dudek:2010wm,Dudek:2011tt,Dudek:2011bn,Wang:2023whb}. 

Recently, a new hadron configuration, the double-gluon hybrid meson consisting of a pair of quark-antiquark and two valence gluons ($\bar{q}GGq$), has been firstly proposed and investigated in the light quark sector~\cite{Chen:2021smz,Su:2022fqr,Su:2023jxb} and then extended to the heavy quark sector~\cite{Tang:2021zti,Su:2023aif}. It is no doubt that one more valence gluon field will enrich the spectra of the hybrid mesons, motivated by recent investigations of the two-gluon and three-gluon glueballs~\cite{Chen:2021cjr,Chen:2021bck}. In ref.~\cite{Tang:2021zti}, the heavy quarkonium double-gluon hybrid mesons with $J^{PC}=1^{--}$ were studied by using the vector interpolating currents with antisymmetric glueball operators. However, an incorrect gluon propagator was adopted in their calculations, while the quantum numbers of two of the hybrid operators were proposed in a wrong {\it C}-parity. In ref.~\cite{Su:2023aif}, the masses and strong decays of charmonium double-gluon hybrids were investigated by using the interpolating currents with an even number of Lorentz indices, leading to very complicated tensor structures of the correlation functions, especially for high-spin states. 

In this work, we shall revisit the heavy quarkonium double-gluon hybrid mesons with exotic quantum numbers $J^{PC}=1^{-+}$ and $2^{+-}$ in the framework of QCD sum rule, which has been proven to be a very powerful non-perturbative approach for studying hadron properties~\cite{Shifman:1978bx,Reinders1985,Colangelo2000,Narison2007,Gubler:2018ctz}. It is a QCD based phenomenological method investigating the bound state problem from the asymptotic freedom side at short distances to long distances where confinement emerges and hadrons are formed. We shall construct distinct interpolating currents with fewer Lorentz indices to simplify the tensor structures of correlation functions. Moreover, it is found that one kind of additional Feynman diagrams with non-local divergences were neglected in previous works~\cite{Chen:2021smz,Su:2023jxb,Su:2023aif,Su:2022fqr,Tang:2021zti}, which will be important after the operator renormalization for the interpolating currents with antisymmetric glueball operators. They cannot be neglected for the correlation functions in which the dimension-4 gluon condensate vanishes.

The rest of this paper is organized as follows. In section~\ref{sec2}, we study the color structures of the $\bar{Q}GGQ$ double-gluon hybrid operator and construct the interpolating currents carrying $J^{PC}=1^{-+}, 2^{+-}$ with one, two and three Lorentz indices. In section~\ref{sec3}, the two-point correlation functions and spectral functions are calculated. In section~\ref{sec4}, we perform  numerical analyses and extract hadron masses of the charmonium and bottomonium double-gluon hybrids. The last section is a brief summary.

%===========================================================
\section{Double-gluon hybrid meson interpolating currents}\label{sec2}
%===========================================================
Following refs.~\cite{Chen:2021smz,Su:2022fqr,Su:2023jxb,Tang:2021zti,Su:2023aif}, we firstly construct the interpolating currents for the heavy quarkonium double-gluon hybrid mesons with exotic quantum numbers $J^{PC}=1^{-+}$ and $2^{+-}$. A double-gluon hybrid meson operator $\bar{Q}GGQ$ is composed of a quark-antiquark $\bar{Q}Q$ field and a two-gluon glueball $GG$ field. In the color SU(3) symmetry, the color structures for the $\bar{Q}GGQ$ operator can be obtained as
\begin{equation}
\begin{aligned}
(\overline{\mathbf{3}} \otimes \mathbf{3})_{[\bar{Q}Q]} \otimes(\mathbf{8} \otimes \mathbf{8})_{[GG]} &=(\mathbf{1} \oplus \mathbf{8})_{[\bar{Q}Q]} \otimes(\mathbf{1} \oplus \mathbf{8} \oplus \mathbf{8} \oplus \mathbf{10} \oplus \overline{\mathbf{1 0}} \oplus \mathbf{27})_{[GG]} \\
&=(\mathbf{1} \otimes \mathbf{1}) \oplus(\mathbf{8} \otimes \mathbf{8})\oplus(\mathbf{8} \otimes \mathbf{8})\oplus\cdots\,,
\end{aligned}
\end{equation}
in which the color singlet structure may come from the $(\mathbf{1}_{[\bar{Q}Q]} \otimes \mathbf{1}_{[GG]})$ and $(\mathbf{8}_{[\bar{Q}Q]} \otimes \mathbf{8}_{[GG]})$ terms. Neglecting the Lorentz structure, a color singlet double-gluon hybrid meson operator has the following form
\begin{equation}\label{eq: operatorform1}
    \bar{Q}_{a}(T^{s}T^{t})_{ab}Q_{b}G^{s}G^{t}\, ,
\end{equation}
where $Q=c/b$ represents a charm/bottom quark field, $G^{s}$ is the gluon field strength, $a, b$ are the color indices, $T^{s}=\lambda^s/2$ ($s=1, 2,...,8$) are the generators of SU(3) group. The generators satisfy the following relation
\begin{equation}\label{Trelation}
    T^{s}T^{t}=\frac{1}{2}\left[\frac{1}{3}\delta^{st}\mathds{1}_{3\times3} +(d^{rst}+i f^{rst})T^{r}\right]\, ,
\end{equation}
where $\mathds{1}_{3\times3}$ is the three-dimension identity matrix, $f^{rst}$ and $d^{rst}$ are the completely antisymmetric structure constants and completely symmetric constants of SU(3) group respectively. Using this relation, the double-gluon hybrid meson operator in Eq.~\eqref{eq: operatorform1} can be separated into three terms with different color structures
\begin{equation}\label{eq: operatorform2}
    \bar{Q}_{a}(T^{s}T^{t})_{ab}Q_{b}G^{s}G^{t}=\frac{1}{6}\bar{Q}_{a}Q_{a}G^{t}G^{t}+\frac{1}{2}\bar{Q}_{a}(T^{r})_{ab}Q_{b}d^{rst}G^{s}G^{t}+\frac{i}{2}\bar{Q}_{a}(T^{r})_{ab}Q_{b}f^{rst}G^{s}G^{t}\, .
\end{equation}
In the above decomposition, the first term is composed of two color singlet field in the color structure $(\mathbf{1}_{[\bar{Q}Q]} \otimes \mathbf{1}_{[GG]})$,  which may couple to the meson-glueball molecule~\cite{Petrov:2022ipv}. The second and third terms are in the octet-octet $(\mathbf{8}_{[\bar{Q}Q]} \otimes \mathbf{8}_{[GG]})$ color structure,  containing symmetric and antisymmetric two-gluon glueball operators, respectively. In this work, we shall study the heavy quarkonium double-gluon hybrid mesons by constructing the interpolating currents in the octet-octet $(\mathbf{8}_{[\bar{Q}Q]} \otimes \mathbf{8}_{[GG]})$ color structure containing both symmetric and antisymmetric glueball operators. 
\begin{table}[htbp]
    \centering
    \renewcommand\arraystretch{1.5}
    \setlength{\tabcolsep}{6mm}
    {
    \begin{tabular}{ccc}
        \hline\hline
     Operator   &  $P$   &$C$\\ \hline
        $\bar{Q}_{a}Q_{b}$ &  $+$   &   $+$\\
        $\bar{Q}_{a}\gamma_{5}Q_{b}$   &  $-$   &$+$\\
        $\bar{Q}_{a}\gamma_{\mu}Q_{b}$   &  $(-1)^{\mu}$   &$-$\\
        $\bar{Q}_{a}\gamma_{\mu}\gamma_{5}Q_{b}$   &  $-(-1)^{\mu}$   &$+$\\
        $\bar{Q}_{a}\sigma_{\mu\nu}Q_{b}$   &  $(-1)^{\mu}(-1)^{\nu}$   &$-$\\
                $d^{rst}G^{s}_{\alpha\beta}G^{t}_{\gamma\delta}T^{r}$ &  $(-1)^{\alpha}(-1)^{\beta}(-1)^{\gamma}(-1)^{\delta}$   &   $+$\\
        $f^{rst}G^{s}_{\alpha\beta}G^{t}_{\gamma\delta}T^{r}$ &  $(-1)^{\alpha}(-1)^{\beta}(-1)^{\gamma}(-1)^{\delta}$   &   $-$\\
        $d^{rst}\widetilde{G}^{s}_{\alpha\beta}G^{t}_{\gamma\delta}T^{r}$ &  $-(-1)^{\alpha}(-1)^{\beta}(-1)^{\gamma}(-1)^{\delta}$   &   $+$\\
        $f^{rst}\widetilde{G}^{s}_{\alpha\beta}G^{t}_{\gamma\delta}T^{r}$ &  $-(-1)^{\alpha}(-1)^{\beta}(-1)^{\gamma}(-1)^{\delta}$   &   $-$\\
        \hline \hline
    \end{tabular}
    }
    \caption{Parity and {\it C}-parity of the color octet quark-antiquark and double-gluon operators.}
\label{table:PC_qq+gg}
\end{table}

Now let's consider the Lorentz structures of these double-gluon hybrid meson operators to determine their $J^{PC}$ quantum numbers. 
We summarize the parities and {\it C}-parities of the color octet quark-antiquark $\bar{Q}Q$ and two-gluon $GG$ operators with various Lorentz structures in table~\ref{table:PC_qq+gg}. Combining these $\bar{Q}Q$ and $GG$ operators, we can construct the interpolating currents for the heavy quarkonium double-gluon hybrid mesons with exotic quantum numbers $J^{PC}=1^{-+}$ and $2^{+-}$ as the following. 
\begin{itemize}
    \item There are six double-gluon hybrid meson interpolating currents with one Lorentz index carrying exotic quantum numbers $J^{PC}=1^{-+}$:
    \begin{equation}\label{eq: currentoperator1}
        \begin{aligned}
            J_{\mu}^{1}&=\bar{Q}_{a}T^{r}_{ab}\gamma_{\rho}Q_{b}g_s^2f^{rst}G^{s}_{\mu\nu}G^{t,\,\nu\rho}\,,\\
             J_{\mu}^{2}&=\bar{Q}_{a}T^{r}_{ab}\gamma_{\rho}Q_{b}g_s^2f^{rst}\widetilde{G}^{s}_{\mu\nu}\widetilde{G}^{t,\,\nu\rho}\,, \\
             J_{\mu}^{3}&=\bar{Q}_{a}T^{r}_{ab}\gamma_{\mu}Q_{b}g_s^2f^{rst}G^{s}_{\alpha\beta}G^{t,\,\alpha\beta}\,,\\
            J_{\mu}^{4}&=\bar{Q}_{a}T^{r}_{ab}\gamma_{\mu}\gamma_{5}Q_{b}g_s^2d^{rst}\widetilde{G}^{s}_{\alpha\beta}G^{t,\,\alpha\beta}\,, \\
            J_{\mu}^{5}&=\bar{Q}_{a}T^{r}_{ab}\gamma_{\rho}\gamma_{5}Q_{b}g_s^2d^{rst}\widetilde{G}^{s}_{\mu\nu}G^{t,\,\nu\rho}\,, \\
            J_{\mu}^{6}&=\bar{Q}_{a}T^{r}_{ab}\gamma_{\rho}\gamma_{5}Q_{b}g_s^2d^{rst}G^{s}_{\mu\nu}\widetilde{G}^{t,\,\nu\rho}\, ,
        \end{aligned}
    \end{equation} 
where $g_s$ is the strong coupling and $\widetilde{G}^{s}_{\mu\nu}=\frac{1}{2}\varepsilon _{\mu\nu\alpha\beta}G^{s,\,\alpha\beta}$ is the dual of gluon field strength. However, one can demonstrate that these six currents are not independent due to the symmetries of the glueball operators. Since $\widetilde{G}^{s}_{\mu\nu}=\frac{1}{2}\varepsilon _{\mu\nu\alpha\beta}G^{s,\,\alpha\beta}$ and $G^{s}_{\mu\nu}=-\frac{1}{2}\varepsilon _{\mu\nu\alpha\beta}\widetilde{G}^{s,\,\alpha\beta}$, we find two relations as 
\begin{equation}\label{eq: relationglueball}
    \begin{aligned}
        \widetilde{G}^{s}_{\mu\nu}G^{t,\,\nu\rho}&=-\frac{1}{4}\varepsilon_{\mu\nu\gamma\lambda}\varepsilon^{\nu\rho\alpha\beta}G^{s,\,\gamma\lambda}\widetilde{G}^{t}_{\alpha\beta}=-\frac{1}{4}(4G^{s,\,\nu\rho}\widetilde{G}^{t}_{\mu\nu}+2\delta^{\,\,\rho}_{\mu}G^{s,\,\alpha\beta}\widetilde{G}^{t}_{\alpha\beta})\\
        &=-G^{s,\,\nu\rho}\widetilde{G}^{t}_{\mu\nu}-\frac{1}{2}\delta^{\,\,\rho}_{\mu}G^{s,\,\alpha\beta}\widetilde{G}^{t}_{\alpha\beta}\,,\\
        \widetilde{G}^{s}_{\mu\nu}\widetilde{G}^{t,\,\nu\rho}&=\frac{1}{4}\varepsilon_{\mu\nu\gamma\lambda}\varepsilon^{\nu\rho\alpha\beta}G^{s,\,\gamma\lambda}G^{t}_{\alpha\beta}=\frac{1}{4}(4G^{s,\,\nu\rho}G^{t}_{\mu\nu}+2\delta^{\,\,\rho}_{\mu}G^{s,\,\alpha\beta}G^{t}_{\alpha\beta})\\
        &=G^{s,\,\nu\rho}G^{t}_{\mu\nu}+\frac{1}{2}\delta^{\,\,\rho}_{\mu}G^{s,\,\alpha\beta}G^{t}_{\alpha\beta}\,,
    \end{aligned}
\end{equation}
in which we have performed the well-known identity
\begin{equation}\label{eq: levicivitaidentity}
    \varepsilon_{\mu\nu\gamma\lambda}\varepsilon^{\nu\rho\alpha\beta}=
    \begin{vmatrix}
        \delta^\rho_\mu&\delta^\rho_\gamma&\delta^\rho_\lambda\\
        \delta^\alpha_\mu&\delta^\alpha_\gamma&\delta^\alpha_\lambda\\
        \delta^\beta_\mu&\delta^\beta_\gamma&\delta^\beta_\lambda
    \end{vmatrix}.
\end{equation}
We find the following relations by using the results in Eq.~\eqref{eq: relationglueball}
\begin{subequations}\label{eq: glueballrelation}
    \begin{align}
        f^{rst}G^{s}_{\alpha\beta}G^{t,\,\alpha\beta}&=-f^{rst}G^{t}_{\alpha\beta}G^{s,\,\alpha\beta}=-f^{rst}G^{s}_{\alpha\beta}G^{t,\,\alpha\beta}\Rightarrow f^{rst}G^{s}_{\alpha\beta}G^{t,\,\alpha\beta}=0\,,\label{eq: glueballrelation1} \\
        f^{rst}\widetilde{G}^{s}_{\mu\nu}\widetilde{G}^{t,\,\nu\rho}&=f^{rst}(G^{s,\,\nu\rho}G^{t}_{\mu\nu}+\frac{1}{2}\delta^{\,\,\rho}_{\mu}G^{s,\,\alpha\beta}G^{t}_{\alpha\beta})=f^{rst}G^{s,\,\nu\rho}G^{t}_{\mu\nu}=-f^{rst}G^{s}_{\mu\nu}G^{t,\,\nu\rho}\,,\label{eq: glueballrelation2}\\
        d^{rst}\widetilde{G}^{s}_{\mu\nu}G^{t,\,\nu\rho}&=d^{rst}\widetilde{G}^{t}_{\mu\nu}G^{s,\,\nu\rho}=-d^{rst}(G^{s,\,\nu\rho}\widetilde{G}^{t}_{\mu\nu}+\frac{1}{2}\delta^{\,\,\rho}_{\mu}G^{s,\,\alpha\beta}\widetilde{G}^{t}_{\alpha\beta})\nonumber\\
        &\Rightarrow d^{rst}\widetilde{G}^{s}_{\mu\nu}G^{t,\,\nu\rho}=d^{rst}G^{s}_{\mu\nu}\widetilde{G}^{t,\,\nu\rho}=-\frac{1}{4}\delta^{\,\,\rho}_{\mu}d^{rst}G^{s,\,\alpha\beta}\widetilde{G}^{t}_{\alpha\beta}\,.\label{eq: glueballrelation3}
        \end{align}
\end{subequations}
So that the six currents in Eqs.~\eqref{eq: currentoperator1} are not independent 
\begin{align}\label{eq: relation1-+}
    J_{\mu}^{1}=-J_{\mu}^{2},\quad J_{\mu}^{3}=0,\quad J_{\mu}^{5}=J_{\mu}^{6}=-\frac{1}{4}J_{\mu}^{4}.
\end{align}
We shall use the interpolating currents $J_{\mu}^{1}$ and $J_{\mu}^{5}$ in the following calculations to study the heavy quarkonium double-gluon hybrid mesons with $J^{PC}=1^{-+}$. Noting that the currents $J_{\mu}^{1}$ and $J_{\mu}^{2}$ were first constructed and studied in ref.~\cite{Tang:2021zti}, in which their quantum numbers were considered incorrectly as $J^{PC}=1^{--}$. Moreover, the calculations in ref.~\cite{Tang:2021zti} are in contradiction to the relation $J_{\mu}^{1}=-J_{\mu}^{2}$ in Eq.~\eqref{eq: relation1-+}. In this work, we shall reexamine these two currents and our calculations of the two-point correlation functions in appendix~\ref{appendix: correlation_spectral} can satisfy exactly the relation $J_{\mu}^{1}=-J_{\mu}^{2}$.
    \item There is only one double-gluon hybrid meson interpolating current with two Lorentz indices carrying exotic quantum numbers $J^{PC}=2^{+-}$:
    \begin{equation}\label{eq: currentoperator2}
        \begin{aligned}
  J_{\mu\nu}^{1}&=\bar{Q}_{a}T^{r}_{ab}\gamma_{5}Q_{b}g_s^2f^{rst}\widetilde{G}^{s}_{\mu\alpha}G^{t,\,\alpha}_{\nu}\, .
        \end{aligned}
    \end{equation}
As shown in the appendix~\ref{appendix: correlation_spectral}, however, the two-point correlation function induced by $J_{\mu\nu}^{1}$ contributes only to the perturbation theory, while the non-perturbative effects up to dimension-eight condensate vanish at the leading order of $\alpha_s$. The OPE series of the current $J_{\mu\nu}^{1}$ are too simple to study the heavy quarkonium double-gluon hybrid mesons reliably.
Revisiting the $\bar{Q}Q$ and $GG$ operators in table~\ref{table:PC_qq+gg}, one may wonder whether the following two $\bar{Q}GGQ$ currents with two Lorentz indices can carry the exotic quantum numbers $J^{PC}=2^{+-}$
    \begin{equation}
        \begin{aligned}
            J_{\mu\nu}^{2}&=\bar{Q}_{a}T^{r}_{ab}Q_{b}g_s^2f^{rst}G^{s}_{\mu\alpha}G^{t,\,\alpha}_{\nu}\,,\\
              J_{\mu\nu}^{3}&=\bar{Q}_{a}T^{r}_{ab}\sigma^{\alpha\beta}Q_{b}g_s^2d^{rst}G^{s}_{\mu\alpha}G^{t}_{\beta\nu}\, . 
        \end{aligned}
    \end{equation}
Unfortunately, it is shown that $J_{\mu\nu}^{2}$ and $J_{\mu\nu}^{3}$ are antisymmetric on the Lorentz indices considering the properties of $d^{rst}$ and $f^{rst}$, so that they cannot carry the tensor quantum numbers $J^{PC}=2^{+-}$. To investigate the $2^{+-}$ heavy quarkonium double-gluon hybrid mesons in QCD sum rules, we need to invoke the interpolating currents with three Lorentz indices, as our previous work for the one-gluon hybrid mesons in ref.~\cite{Wang:2023whb}. 
    \item There are four double-gluon hybrid meson interpolating currents with three Lorentz indices carrying exotic quantum numbers $J^{PC}=2^{+-}$:
    \begin{equation}\label{eq:currentoperator3}
        \begin{aligned}
            J_{\mu\nu\rho}^{1}&=\bar{Q}_{a}T^{r}_{ab}\gamma_{\alpha}Q_{b}g_s^2d^{rst}G^{s}_{\mu\nu}G^{t,\,\alpha}_{\rho}\,,\\
              J_{\mu\nu\rho}^{2}&=\bar{Q}_{a}T^{r}_{ab}\gamma_{\alpha}Q_{b}g_s^2d^{rst}G^{s}_{\mu\nu}\widetilde{G}^{t,\,\alpha}_{\rho}\,,\\
            J_{\mu\nu\rho}^{3}&=\bar{Q}_{a}T^{r}_{ab}\gamma_{\alpha}\gamma_{5}Q_{b}g_s^2f^{rst}G^{s}_{\mu\nu}G^{t,\,\alpha}_{\rho}\,,\\
           J_{\mu\nu\rho}^{4}&=\bar{Q}_{a}T^{r}_{ab}\gamma_{\alpha}\gamma_{5}Q_{b}g_s^2f^{rst}G^{s}_{\mu\nu}\widetilde{G}^{t,\,\alpha}_{\rho}\, .
        \end{aligned}
    \end{equation}
One can obtain their dual currents $\tilde{J}_{\mu\nu\rho}^{1}, \tilde{J}_{\mu\nu\rho}^{2}, \tilde{J}_{\mu\nu\rho}^{3}, \tilde{J}_{\mu\nu\rho}^{4}$ by replacing $G^{s}_{\mu\nu}$ with $\widetilde{G}^{s}_{\mu\nu}$. However, a current $J_{\mu\nu\rho}^{i}$ and its dual $\tilde{J}_{\mu\nu\rho}^{i}$ actually couple to the same $2^{+-}$ hybrid state via different coupling tensor structures, as shown in ref.~\cite{Wang:2023whb}. 
\end{itemize}

In this work, we shall use the interpolating currents $J_{\mu}^{1}, J_{\mu}^{5}$ in Eq.~\eqref{eq: currentoperator1} to investigate the $1^{-+}$ heavy quarkonium double-gluon hybrid mesons, while the currents $J_{\mu\nu\rho}^{1}, J_{\mu\nu\rho}^{2}, J_{\mu\nu\rho}^{3}, J_{\mu\nu\rho}^{4}$ in Eq.~\eqref{eq:currentoperator3} to study the $2^{+-}$ hybrid mesons.
%================================================================================
\section{QCD sum rules for the heavy quarkonium double-gluon hybrids}\label{sec3}
%================================================================================
The two-point correlation functions induced by the interpolating currents with one, two and three Lorentz indices are defined as
\begin{eqnarray}
\label{correlation1}
\Pi_{\mu\nu }(p^2)&=&i\int d^4x e^{ip\cdot x}\left\langle 0\left|T\left[J_{\mu}(x) J^\dagger_{\nu }(0)\right]\right|0\right\rangle\, , \\
\label{correlation2}
\Pi_{\mu_1\nu_1,\,\mu_2\nu_2}(p^2)&=&i\int d^4x e^{ip\cdot x}\left\langle 0\left|T\left[J_{\mu_1\nu_1}(x) J^\dagger_{\mu_2\nu_2}(0)\right]\right|0\right\rangle\, ,\\
\label{correlation3}
\Pi_{\mu_1\nu_1\rho_1,\,\mu_2\nu_2\rho_2}(p^{2}) &=& i \int d^{4} x e^{i p \cdot x}\left\langle 0\left|T\left[J_{\mu_1\nu_1\rho_1}(x) J_{\mu_2\nu_2\rho_2}^{\dagger}(0)\right]\right| 0\right\rangle\, .
\end{eqnarray}
In general, these correlation functions contain several different invariant functions contributing to hadron states with various quantum numbers. To investigate the interested hadron state, one needs to extract the corresponding invariant function from the above correlation functions. 
A current with one Lorentz index can couple to hadron states with spin-1 and spin-0 via
\begin{subequations}\label{eq: couplingrelation1}
    \begin{align}
        \left\langle 0 \left\lvert J_{\mu}\right\rvert 1^{P}(p)\right\rangle &=Z^1_{1}\epsilon_{\mu}\,,\\
        \left\langle 0 \left\lvert J_{\mu}\right\rvert 0^{(-P)}(p)\right\rangle &=Z^0_{1}p_{\mu}\,,
    \end{align}
\end{subequations}
where $\epsilon_{\mu}$ is the spin-1 polarization vector. And a current with two Lorentz indices can couple to hadron states with spin-2, spin-1 and spin-0 via the coupling relations 
\begin{subequations}\label{eq: couplingrelation2}
    \begin{align}
        \left\langle 0 \left\lvert J_{\mu\nu}\right\rvert 2^{P}(p)\right\rangle &=Z^2_{1}\epsilon_{\mu\nu}\,,\\
        \left\langle 0 \left\lvert J_{\mu\nu}\right\rvert 1^{P}(p)\right\rangle &=Z^1_{1}\varepsilon_{\mu\nu\alpha\beta}\epsilon^{\alpha}p^{\beta}\,,\\
        \left\langle 0 \left\lvert J_{\mu\nu}\right\rvert 1^{(-P)}(p)\right\rangle &=Z^1_{2}\epsilon_{\mu}p_{\nu}+Z^1_{3}\epsilon_{\nu}p_{\mu}\,,\\
        \left\langle 0 \left\lvert J_{\mu\nu}\right\rvert 0^{P}(p)\right\rangle &=Z^0_{1}g_{\mu\nu}+Z^0_{2}p_{\mu}p_{\nu}\,,
    \end{align}
\end{subequations}
where $\epsilon_{\mu\nu}$ is the symmetric spin-2 polarization tensor. Likewise, a current with three Lorentz indices can couple to hadron states with spin-3, spin-2, spin-1 and spin-0 via the following coupling relations~\cite{Wang:2023whb}
\begin{subequations}\label{eq: couplingrelation3}
    \begin{align}
       \left\langle 0\left|J_{\mu \nu \rho}\right| 3^{P}(p)\right\rangle&=Z_1^3 \epsilon_{\mu \nu \rho} \, ,\\
        \left\langle 0\left|J_{\mu \nu \rho}\right| 2^{(-P)}(p)\right\rangle&=Z_1^2 \epsilon_{\mu \nu} p_\rho+Z_2^2 \epsilon_{\mu \rho} p_\nu+Z_3^2 \epsilon_{\nu \rho} p_\mu  \,,\\
        \left\langle 0\left|J_{\mu \nu \rho}\right| 2^{P}(p)\right\rangle&=Z_4^2 \varepsilon_{\mu \nu\tau \theta} \epsilon_\rho^{~\tau} p^\theta+Z_5^2 \varepsilon_{\mu \rho \tau \theta} \epsilon_\nu^{~\tau} p^\theta  \,,\\
         \left\langle 0\left|J_{\mu \nu \rho}\right| 1^{P}(p)\right\rangle&=Z_1^1 \epsilon_\mu g_{\nu \rho}+Z_2^1 \epsilon_\nu g_{\mu \rho}+Z_3^1 \epsilon_\rho g_{\mu \nu}+Z_4^1 \epsilon_\mu p_\nu p_\rho+Z_5^1 \epsilon_\nu p_\mu p_\rho+Z_6^1 \epsilon_\rho p_\mu p_\nu  \,,\\
        \left\langle 0\left|J_{\mu \nu \rho}\right| 1^{(-P)}(p)\right\rangle&=Z_7^1 \varepsilon_{\mu \nu \rho\tau} \epsilon^\tau+Z_8^1 \varepsilon_{\mu \nu \tau \lambda} \epsilon^\tau p^\lambda p_\rho+Z_9^1 \varepsilon_{\mu \rho \tau \lambda} \epsilon^\tau p^\lambda p_\nu  \,,\\
        \left\langle 0\left|J_{\mu \nu \rho}\right| 0^{(-P)}(p)\right\rangle&=Z_1^0 p_\mu g_{\nu \rho}+Z_2^0 p_\nu g_{\mu \rho}+Z_3^0 p_\rho g_{\mu \nu}+Z_4^0 p_\mu p_\nu p_\rho  \,,\\
        \left\langle 0\left|J_{\mu \nu \rho}\right| 0^{P}(p)\right\rangle&=Z_5^0 \varepsilon_{\mu \nu \rho\tau} p^\tau  \,,
    \end{align}
\end{subequations}
where $\epsilon_{\mu\nu\rho}$ is the spin-3 polarization tensor. One notes that all the double-gluon hybrid interpolating currents in Eq.~\eqref{eq:currentoperator3} cannot couple to the spin-3 states because the first two Lorentz indices in these currents are antisymmetric while the spin-3 polarization tensor $\epsilon_{\mu\nu\rho}$ is completely symmetric. Using the above coupling relations, the two-point correlation functions defined in Eqs.~\eqref{correlation1}-\eqref{correlation3} can be written as ~\cite{Govaerts:1986pp,Chen:2013zia,Wang:2023whb}
\begin{align}
    \Pi_{\mu\nu}(p^2)&=\eta_{\mu\nu}\Pi^{1^{P}}(p^2)+\frac{p_{\mu}p_{\nu}}{p^2}\Pi^{0^{(-P)}}(p^2)\,,\label{eq: tensorstru1}\\
    \Pi_{\mu_{1}\nu_{1},\, \mu_{2}\nu_{2}}(p^2)&=\frac{1}{2}(\eta_{\mu_{1}\mu_{2}}\eta_{\nu_{1}\nu_{2}}+\eta_{\mu_{1}\nu_{2}}\eta_{\nu_{1}\mu_{2}}-\frac{2}{3}\eta_{\mu_{1}\nu_{1}}\eta_{\mu_{2}\nu_{2}})\Pi^{2^{P}}(p^2)+\cdots\,,\label{eq: tensorstru2}\\
    \Pi_{\mu_{1}\nu_{1}\rho_{1},\, \mu_{2}\nu_{2}\rho_{2}}(p^2)&=\mathbb{P}^{2^{P}}_{\mu_{1}\nu_{1}\rho_{1},\, \mu_{2}\nu_{2}\rho_{2}}\Pi^{2^{P}}(p^2)+\mathbb{P}^{2^{(-P)}}_{\mu_{1}\nu_{1}\rho_{1},\, \mu_{2}\nu_{2}\rho_{2}}\Pi^{2^{(-P)}}(p^2)+\cdots\,,\label{eq: tensorstru3}
\end{align}
in which $\Pi^{0^{(-P)}}(p^2), \Pi^{1^{P}}(p^2), \Pi^{2^{P}}(p^2)$ and $\Pi^{2^{(-P)}}(p^2)$ are the invariant functions contributing to the hadron states with spin-parity $J^P=0^{(-P)}, 1^P, 2^P$ and $2^{(-P)}$, respectively. In Eqs.~\eqref{eq: tensorstru2}-\eqref{eq: tensorstru3}, there are also contributions from other invariant functions in ``$\cdots$'', which are omitted here because we don't concern these structures in this work. The tensor projectors $\eta_{\mu\nu}$, $\mathbb{P}^{2^{P}}_{\mu_{1}\nu_{1}\rho_{1},\, \mu_{2}\nu_{2}\rho_{2}}$ and $\mathbb{P}^{2^{(-P)}}_{\mu_{1}\nu_{1}\rho_{1},\, \mu_{2}\nu_{2}\rho_{2}}$ in Eqs.~\eqref{eq: tensorstru1}-\eqref{eq: tensorstru3} are defined as
\begin{align}
    \eta_{\mu\nu}&=-g_{\mu\nu}+\frac{p_{\mu}p_{\nu}}{p^2}\,,\\
    \mathbb{P}^{2^{P}}_{\mu_{1}\nu_{1}\rho_{1},\, \mu_{2}\nu_{2}\rho_{2}}&=\varepsilon_{\mu_{1}\nu_{1}\tau_{1}\theta_{1}}\varepsilon_{\mu_{2}\nu_{2}\tau_{2}\theta_{2}}\frac{p^{\theta_{1}}p^{\theta_{2}}}{p^2}\sum \epsilon^{~\tau_{1}}_{\rho_{1}} \epsilon^{~\tau_{2}\ast }_{\rho_{2}}\,,\\
    \mathbb{P}^{2^{(-P)}}_{\mu_{1}\nu_{1}\rho_{1},\, \mu_{2}\nu_{2}\rho_{2}}&=\frac{1}{p^2}\sum (\epsilon_{\mu_{1}\rho_{1}}p_{\nu_{1}}-\epsilon_{\nu_{1}\rho_{1}}p_{\mu_{1}})(\epsilon^{\ast}_{\mu_{2}\rho_{2}}p_{\nu_{2}}-\epsilon^{\ast}_{\nu_{2}\rho_{2}}p_{\mu_{2}})\,,
\end{align}
with the summation over the polarization tensor  
\begin{equation}\label{eq: sumpolar}
    \sum \epsilon_{\mu_{1}\nu_{1}}\epsilon_{\mu_{2}\nu_{2}}^{\ast}=\frac{1}{2}(\eta_{\mu_{1}\mu_{2}}\eta_{\nu_{1}\nu_{2}}+\eta_{\mu_{1}\nu_{2}}\eta_{\nu_{1}\mu_{2}}-\frac{2}{3}\eta_{\mu_{1}\nu_{1}}\eta_{\mu_{2}\nu_{2}})\,.
\end{equation}

At the hadron level, the invariant function can be described by the dispersion relation 
\begin{equation}
\Pi\left(p^{2}\right)=\frac{\left(p^{2}\right)^{N}}{\pi} \int_{4m_{Q}^{2}}^{\infty} \frac{\operatorname{Im} \Pi(s)}{s^{N}\left(s-p^{2}-i \epsilon\right)} d s+\sum_{n=0}^{N-1} b_{n}\left(p^{2}\right)^{n}\, ,
\label{Cor-Spe}
\end{equation}
where $b_n$ are the subtraction constants and will be removed later by performing the Borel transform. The imaginary part of the invariant function is usually defined as the the spectral function 
\begin{equation}\label{eq: rhos}
    \rho(s)\equiv\frac{1}{\pi}\text{Im} \Pi(s)=\sum_{n}\delta(s-m_n^2)\langle 0\lvert J\rvert n\rangle \langle n \lvert J^{\dagger} \rvert 0 \rangle  =f_X^2\delta(s-m_X^2)+\cdots,
\end{equation}
in which the ``one pole dominant narrow resonance'' approximation is adopted in the last step, and ``$\cdots$'' represents the contributions from the  continuum and higher excited states, and $f_X$, $m_X$ are the coupling constant and hadron mass of the lowest lying resonance respectively.

\begin{figure}[htbp]
    \centering
    \subfigure[]{
        \includegraphics[width=0.15\textwidth]{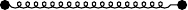}
        \label{fig: gluon_propagator_a}
        }\qquad
    \subfigure[]{
        \includegraphics[width=0.15\textwidth]{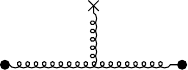}
        \label{fig: gluon_propagator_b}
        }\qquad
    \subfigure[]{
        \includegraphics[width=0.15\textwidth]{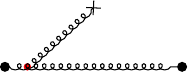}
        \label{fig: gluon_propagator_c}
        }
    \caption{The full gluon propagator of the gluon field strength. The two-gluon-line with a red vertex in (c) represents the single-gluon emission with one gluon leg attached from the third term of Eq.~\eqref{eq: gluon filed strength tensor}.} 
    \label{fig: gluon_propagator}
\end{figure}

At the quark-gluon level, the invariant functions can be calculated via the operator product expansion (OPE) method. In this work, we calculate the correlation functions at the leading order (LO) of $\alpha_s$ and consider the nonperturbative effects of condensates up to dimension eight. In the heavy quarkonium double-gluon hybrid systems, we apply the factorization assumption to reduce the dimension-eight four-gluon condensate as $\left\langle \alpha_{s}GG\right\rangle^2$, which is usually adopted in QCD sum rules to estimate the values of the high dimension condensates~\cite{Shifman:1978bx,Reinders1985}. In our calculations, we use the full heavy quark propagator ($Q=c~\text{or}~b$) in momentum space
\begin{equation}\label{eq: quarkpropagator}
    iS_{ab}(p)=\frac{i\delta_{ab}}{\cancel{p}-m_Q}-\frac{i}{4}g_{s}T^{r}_{ab}G^{r}_{\mu\nu}\frac{\sigma^{\mu\nu}(\cancel{p}+m_Q)+(\cancel{p}+m_Q)\sigma^{\mu\nu}}{(p^2-m_Q^2)^2}\,,
\end{equation}
where the first term is the free quark propagator and the second term is for the single-gluon emission with one gluon leg attached, $m_Q$ is the heavy quark mass. Similarly, the full gluon propagator also contains the free gluon propagator and the single-gluon emission term, as shown in figure~\ref{fig: gluon_propagator_a} and figure~\ref{fig: gluon_propagator_b} respectively. Considering the definition of gluon filed strength tensor 
\begin{equation}\label{eq: gluon filed strength tensor}
    G^{r}_{\mu\nu}=\partial_{\mu}A^{r}_{\nu}-\partial_{\nu}A^{r}_{\mu}+g_s f^{rst}A_{\mu}^{s}A^{t}_{\nu}\,,
\end{equation}
the third term with two gauge fields $A_{\mu}^{s}A^{t}_{\nu}$ will result in  another different single-gluon emission with one gluon leg attached in the full gluon propagator, as shown in figure~\ref{fig: gluon_propagator_c}. 
The expression of the full gluon propagator for the gluon filed strength up to single-gluon emission term in momentum space is 
\begin{equation}\label{eq: gluon_propagator}
    \begin{aligned}
        iS_{\alpha\mu,\, \beta\nu}^{rs}(p)=&\frac{-i\delta^{rs}}{p^2+i\epsilon}\left[g_{\mu\nu}p_{\alpha}p_{\beta}-g_{\mu\beta}p_{\alpha}p_{\nu}+(\mu\leftrightarrow \alpha,~\nu\leftrightarrow \beta)\right]\\
        &+\frac{i g_sf^{rst}}{(p^2+i\epsilon)^2}G^{t,\,\rho\sigma}(0)\bigg[2 p_{\alpha}g_{\mu\rho}(p_{\beta}g_{\nu\sigma}-p_{\nu}g_{\beta\sigma})+p_{\alpha}p_{\sigma}(g_{\mu\beta}g_{\nu\rho}-g_{\mu\nu}g_{\beta\rho})\\
        &\phantom{====\qquad\qquad\qquad}+(\mu\leftrightarrow \alpha,~\nu\leftrightarrow \beta)\bigg]\\
        &+\frac{i g_sf^{rst}}{2(p^2+i\epsilon)^2}G^{t,\,\rho\sigma}(0)\bigg[2 g_{\alpha\sigma}p_{\rho}(g_{\mu\beta}p_{\nu}-g_{\mu\nu}p_{\beta})-p^2g_{\alpha\sigma}(g_{\mu\beta}g_{\nu\rho}-g_{\mu\nu}g_{\beta\rho})\\
        &\phantom{====\qquad\qquad\qquad}+(\mu\leftrightarrow \alpha,~\nu\leftrightarrow \beta)\bigg]\,,
    \end{aligned}
\end{equation}
where the three square brackets correspond to three diagrams in figure~\ref{fig: gluon_propagator}. One notes that this expression is different from that used in ref.~\cite{Tang:2021zti}, leading to different results of the correlation functions. 
%We also need the perturbative propagator of the gluon filed strength tensor in coordinate space~\cite{Govaerts:1984hc}
%\begin{equation}\label{eq: gluon_propagator_pert}
%    iS_{\alpha\mu,\beta\nu}^{rs,\, pert}(x)=\frac{\delta^{rs}}{2\pi^2}\frac{1}{x^6}\left[g_{\mu\nu}(g_{\alpha\beta}x^2-4x_{\alpha}x_{\beta})-g_{\beta\mu}(g_{\alpha\nu}x^2-4x_{\alpha}x_{\nu})+(\mu\leftrightarrow \alpha,~\nu\leftrightarrow \beta)\right].
%\end{equation}

\begin{figure}[t!]
    \centering
    \subfigure[]{
        \includegraphics[width=0.15\textwidth]{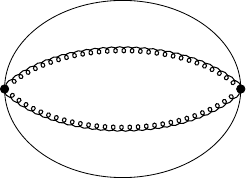}
        \label{fig: feynman_diagram_a}
        }\qquad
    \subfigure[]{
        \includegraphics[width=0.15\textwidth]{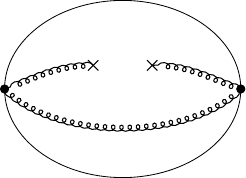}
        \label{fig: feynman_diagram_b}
        }\qquad
        \subfigure[]{
        \includegraphics[width=0.15\textwidth]{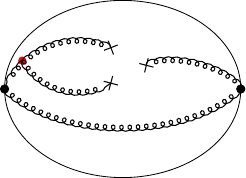}
        \label{fig: feynman_diagram_c}
        }\qquad
        \subfigure[]{
        \includegraphics[width=0.15\textwidth]{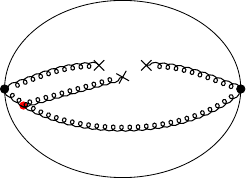}
        \label{fig: feynman_diagram_d}
        }\qquad\\
        \subfigure[]{
        \includegraphics[width=0.15\textwidth]{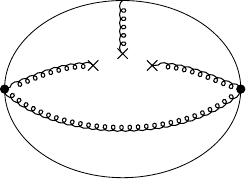}
        \label{fig: feynman_diagram_e}
        }\qquad
        \subfigure[]{
        \includegraphics[width=0.15\textwidth]{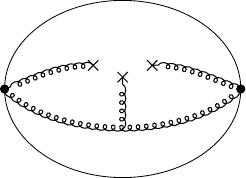}
        \label{fig: feynman_diagram_f}
        }\qquad
        \subfigure[]{
        \includegraphics[width=0.15\textwidth]{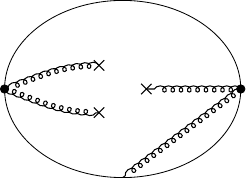}
        \label{fig: feynman_diagram_g}
        }\qquad
        \subfigure[]{
        \includegraphics[width=0.15\textwidth]{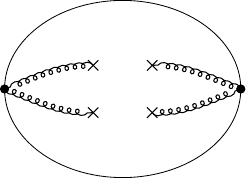}
        \label{fig: feynman_diagram_h}
        }\qquad
        \caption{The LO Feynman diagrams for the $\bar QGGQ$ hybrid meson systems up to dimension eight gluon condensate. Diagrams related by symmetry are not shown.}
        \label{fig: feynman_diagram}
\end{figure}

To calculate the correlation functions at the leading order of $\alpha_s$, we consider the Feynman diagrams depicted in figure~\ref{fig: feynman_diagram}, in which figure~\ref{fig: feynman_diagram_b}-figure~\ref{fig: feynman_diagram_f} are the two-loop sunrise diagrams with one massless line and two equal mass lines, and figure~\ref{fig: feynman_diagram_g} can be reduced to the two-loop sunrise diagram~\cite{Tarasov:1997kx}. For the perturbative diagram in figure~\ref{fig: feynman_diagram_a}, it can also be reduced to a two-loop sunrise diagram by integrating the massless gluon loop. These two-loop sunrise diagrams can be expressed in terms of the generalized hypergeometric functions $_{p+1}F_{p}$~\cite{Broadhurst:1993mw,Bateman:100233}
\begin{equation}\label{eq: sunrise_res}
    \begin{aligned}
        &\int \frac{d^Dk_1}{(2\pi)^D}\frac{d^Dk_2}{(2\pi)^D}\frac{1}{(k_2^2-m_Q^2+i\epsilon)^{\alpha}[(k_1-q)^2-m_Q^2+i\epsilon]^{\beta}[(k_1-k_2)^2+i\epsilon]^{\gamma}}\\
        =&\frac{(-1)^{1-\alpha -\beta -\gamma } \left(m_Q^2\right)^{D-\alpha -\beta -\gamma}}{(4 \pi )^D}\frac{\Gamma \left(\frac{D}{2}-\gamma \right) \Gamma \left(\alpha +\gamma-\frac{D}{2} \right) \Gamma \left(\beta +\gamma -\frac{D}{2}\right) \Gamma (\alpha +\beta +\gamma-D)}{\Gamma (\alpha ) \Gamma (\beta ) \Gamma \left(\frac{D}{2}\right) \Gamma (\alpha +\beta +2 \gamma-D )}\\
        &\times \pFq{4}{3}{\gamma,\alpha +\beta +\gamma-D,\beta +\gamma-\frac{D}{2},\alpha +\gamma-\frac{D}{2}}{\frac{D}{2},\gamma+\frac{\alpha +\beta-D}{2},\gamma+\frac{\alpha +\beta +1-D}{2}}{\frac{q^2}{4 m_Q^2}}\,,
    \end{aligned}
\end{equation}
where the dimensional regularization has been performed with $D=4-2\epsilon$. The one-loop diagrams can also be expressed in terms of hypergeometric functions, as tabulated in refs.~\cite{Boos:1990rg,Davydychev:1990cq}. Using the Mathematica package Tarcer~\cite{Mertig:1998vk} based on the algorithm in refs.~\cite{Tarasov:1996br,Tarasov:1997kx}, the two-loop Feynman integrals with tensor structures can be reduced to the scalar integrals as in Eq.~\eqref{eq: sunrise_res}.

\begin{figure}[htbp]
    \centering
    \subfigure[]{
        \includegraphics[width=0.15\textwidth]{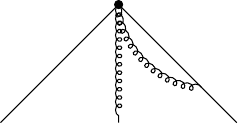}
        \label{fig: counter_diagram_a}
        }\qquad
    \subfigure[]{
        \includegraphics[width=0.15\textwidth]{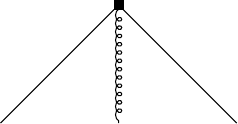}
        \label{fig: counter_diagram_b}
        }\qquad
    \subfigure[]{
        \includegraphics[width=0.15\textwidth]{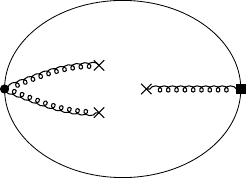}
        \label{fig: counter_diagram_c}
        }\qquad
    \caption{Diagram (a) is a subgraph of figure~\ref{fig: feynman_diagram_g}, raising the non-local divergence exhibited in figure~\ref{fig: feynman_diagram_g}. Diagram (b) with the black square represents the counter term vertex generated from the divergence of diagram (a). Diagram (c) with the counter term vertex will eliminate the non-local divergence in figure~\ref{fig: feynman_diagram_g}.}
    \label{fig: counter_diagram}
\end{figure}

We find that the diagram in figure~\ref{fig: feynman_diagram_g} gives significant contributions to the correlation functions of the interpolating currents $J_{\mu}^{1/2}, J_{\mu\nu\rho}^{3}$ and $J_{\mu\nu\rho}^{4}$ with antisymmetric glueball operator $f^{rst}G^{s}_{\mu\nu}G^{t}_{\alpha\beta}$, while making no contribution to the currents with symmetric glueball operator $d^{rst}G^{s}_{\mu\nu}G^{t}_{\alpha\beta}$. They are especially important for $J_{\mu}^{1/2}$, in which the correlation functions have no dimension-4 condensate contribution. 
The calculations show that there exist non-local divergences in figure~\ref{fig: feynman_diagram_g}, which cannot be removed simply by the subtraction scheme as shown in Eq.~\eqref{Cor-Spe}. To cancel out such non-local divergence in figure~\ref{fig: feynman_diagram_g}, we use the diagrammatic renormalization approach described in refs.~\cite{Collins:1984xc,Muta:1998vi,deOliveira:2022eeq}. We show the renormalization process in figure~\ref{fig: counter_diagram}, in which figure~\ref{fig: counter_diagram_b} represents the counter term vertex introduced to cancel out the divergence arising from figure~\ref{fig: counter_diagram_a}. The vertices of currents $J_{\mu}^{1}, J_{\mu\nu\rho}^{3}, J_{\mu\nu\rho}^{4}$, their respective counter terms, and the corresponding renormalization coefficients are listed in table~\ref{table: renormalization_constant}, appendix~\ref{appendix: renormalization}. In next section, we shall discuss the significance of figure~\ref{fig: feynman_diagram_g} in contributing to the tri-gluon condensate.

To obtain the spectral functions, we need to calculate the imaginary part of the generalized hypergeometric function through its integral representation~\cite{KARP2012348,KARP201742}
\begin{equation}\label{eq: int_rep_hyp}
    \pFq{p+1}{p}{\sigma,A}{B}{z}=\int_{1}^{\infty}\frac{\mu(t)dt}{(t-z)^{\sigma}}\,,
\end{equation}
with
\begin{equation}\label{eq: MeijerGfunc}
    \mu(t)=\left(\prod_{i=1}^{p}\frac{\Gamma(b_{i})}{\Gamma(a_{i})}\right)t^{\sigma-1}\MeijerG{p,0}{p,p}{B}{A}{\frac{1}{t}}\,,
\end{equation}
in which $A=(a_1,a_2,\ldots,a_p)$, $B=(b_1,b_2,\ldots,b_p)$, and $\MeijerG{p,0}{p,p}{B}{A}{\frac{1}{t}}$ is the Meijer's G-function~\cite{Bateman:100233}. In our calculations, we always use the functions with parameter $\sigma=1$ in Eq.~\eqref{eq: int_rep_hyp}, so that the imaginary part of the generalized hypergeometric function can be obtained as
\begin{equation}
    \frac{1}{\pi}\text{Im}\left[\pFq{p+1}{p}{1,A}{B}{z}\right] =\left[\prod_{i=1}^{p}\frac{\Gamma(b_{i})}{\Gamma(a_{i})}\right]\MeijerG{p,0}{p,p}{B}{A}{\frac{1}{z}}\, .
\end{equation}
We collect results of the correlation functions and spectral functions for all interpolating currents in appendix~\ref{appendix: correlation_spectral}. 

The Borel transform is usually applied for the correlation function at both the hadron and quark-gluon levels to improve the OPE convergence and suppress the contributions from continuum and higher excited states. Based on the hadron-quark duality, one can establish the QCD sum rules for the heavy quarkonium double-gluon hybrid  mesons 
\begin{equation}\label{eq: laplace_sr}
    \mathcal{L}_{k}(s_0,\,M_{B}^2)=f_{X}^{2}(m_{X}^{2})^{k}e^{-m_X^2/M_B^2}=\int_{s_<}^{s_0}ds s^{k}\rho_{OPE}(s)e^{-s/M_B^2}\,,
\end{equation}
where $s_0$ is the continuum threshold and $M_B$ is the Borel mass. Then we can extract the lowest lying hadron mass
\begin{equation}\label{eq: mass}
    m_X^2=\frac{\mathcal{L}_{1}(s_0,\,M_{B}^2)}{\mathcal{L}_{0}(s_0,\,M_{B}^2)}\,.
\end{equation}

%================================================================================
\section{Numerical analysis}\label{sec4}
%================================================================================
To perform the QCD sum rule numerical analyses, we use the following values of the heavy quark masses, the strong coupling and the QCD condensates~\cite{ParticleDataGroup:2022pth,Narison:2011xe,Narison:2018dcr}:
\begin{equation}\label{eq: parameter_sr}
    \begin{aligned}
        m_{c}(\mu=\overline{m} _{c})&=(1.27\pm 0.02)~\text{GeV}\,,\\
        \alpha_{s}(\mu=\overline{m}_{c})&=0.38\pm 0.03\,,\\
        m_{b}(\mu=\overline{m}_{b})&=4.18^{+0.04}_{-0.03}~\text{GeV}\,,\\
        \alpha_{s}(\mu=\overline{m}_{b})&=0.223\pm 0.008\,,\\
        \GGb&=(6.35\pm 0.35)\times 10^{-2}~\text{GeV}^4\,,\\
        \GGGb&=(8.2\pm 1)\times \GGb~\text{GeV}^2\,,
    \end{aligned}
\end{equation}
where the heavy quark masses and strong coupling at the corresponding energy scales are obtained by using the two-loop QCD perturbation theory in the $\overline{\text{MS}}$ scheme.

As shown in Eq.~\eqref{eq: mass}, the hadron mass depends on the continuum threshold $s_0$ and Borel mass $M_B$. The suitable working regions of these two parameters are very important to obtain reliable sum rule prediction of the hadron mass. In QCD sum rules, the behaviors of OPE convergence and the pole contribution (PC) are usually studied to determine the parameter working regions. To guarantee the good OPE convergence, we require that the contributions of the condensates with $D=6$ and $D=8$ to be less than $20\%$ and $10\%$ respectively
\begin{equation}\label{eq: convergence}
    \begin{aligned}
        R_{D=6}&=\left\lvert\frac{\mathcal{L}_{0}^{D=6}(\infty,M_B^2)}{\mathcal{L}_{0}^{total}(\infty,M_B^2)}\right\rvert<20\%\,,\\
        R_{D=8}&=\left\lvert\frac{\mathcal{L}_{0}^{D=8}(\infty,M_B^2)}{\mathcal{L}_{0}^{total}(\infty,M_B^2)}\right\rvert<10\%\, ,
    \end{aligned}
\end{equation}
which can give the lower bound on $M_B^2$. 
Furthermore, we require the pole contribution to be larger than $40\%$ to determine the upper bound of $M_B^2$
\begin{equation}\label{eq: PC}
    \text{PC}=\left\lvert\frac{\mathcal{L}_{0}^{total}(s_0,M_B^2)}{\mathcal{L}_{0}^{total}(\infty,M_B^2)}\right\rvert>40\%\,.
\end{equation}
One can find the Borel window of $M_B^2$ with these two requirements, in which an optimal value of the continuum threshold $s_{0}$ can also be determined by minimizing the dependence of the hybrid mass $m_{X}$ on the Borel parameter $M_B^2$. In other words, there should exist a smooth plateau for the hybrid mass $m_{X}$ in the variation of the Borel parameter $M_B^2$.

\begin{figure}[htbp]
    \centering
    \subfigure[]{
        \includegraphics[width=0.45\textwidth]{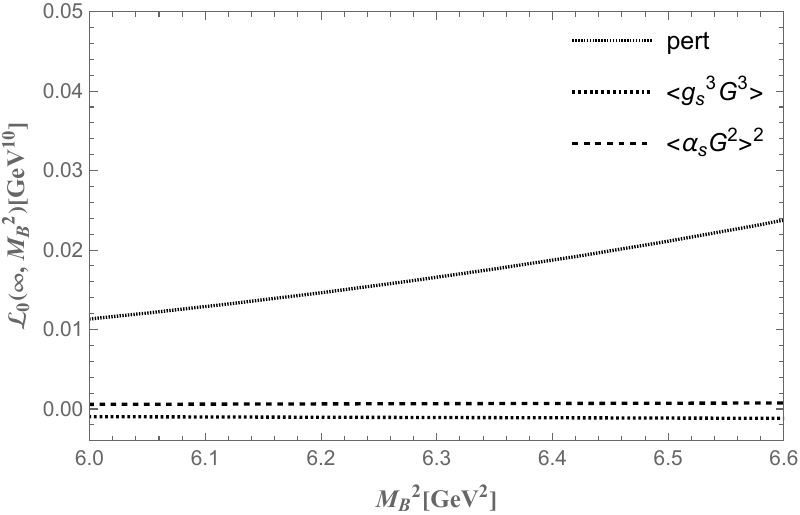}
        \label{fig: ope_J_mu1/2_a}
        }\qquad
    \subfigure[]{
        \includegraphics[width=0.45\textwidth]{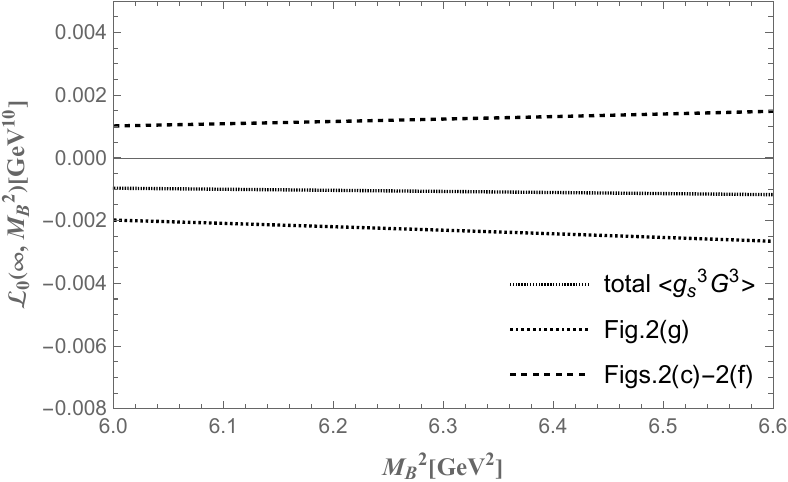}
        \label{fig: ope_J_mu1/2_b}
        }\qquad
    \caption{For the charmonium hybrid system with the interpolating current $J_{\mu}^{1}$ carrying $J^{PC}=1^{-+}$: (a) convergence of the OPE series; (b) important contribution from the Feynman diagram figure~\ref{fig: feynman_diagram_g} to the tri-gluon condensate.}
    \label{fig: ope_J_mu1/2}
\end{figure}

In figure~\ref{fig: ope_J_mu1/2}, we show the OPE series behavior for the  charmonium hybrid system by using the current $J_{\mu}^{1}$, for which the dimension-4 gluon condensate vanishes. One finds that the dominant non-perturbative effect comes from the dimension-6 tri-gluon condensate. We emphasize here the important contribution of the Feynman diagram in figure~\ref{fig: feynman_diagram_g} to the tri-gluon condensate, which is not taken into account in previous studies~\cite{Chen:2021smz,Su:2023jxb,Su:2023aif,Su:2022fqr,Tang:2021zti}. As shown in figure~\ref{fig: ope_J_mu1/2_b}, the tri-gluon condensate contribution from figure~\ref{fig: feynman_diagram_g} is negative while those from Figs.~\ref{fig: feynman_diagram_c}-\ref{fig: feynman_diagram_f} are positive, resulting in the negative value for the tri-gluon condensate in total. Such a big improvement would potentially affect the sum rule stability and mass predictions.
For the interpolating currents $J_{\mu\nu\rho}^{3}$ and $J_{\mu\nu\rho}^{4}$ with $J^{PC}=2^{+-}$, our analyses show similar effect of diagram figure~\ref{fig: feynman_diagram_g} on the tri-gluon condensate, while the influence on the sum rule behaviors is smaller than that for the $J_{\mu}^{1}$ current. 
\begin{figure}[htbp]
    \centering
    \subfigure[]{
        \includegraphics[width=0.45\textwidth]{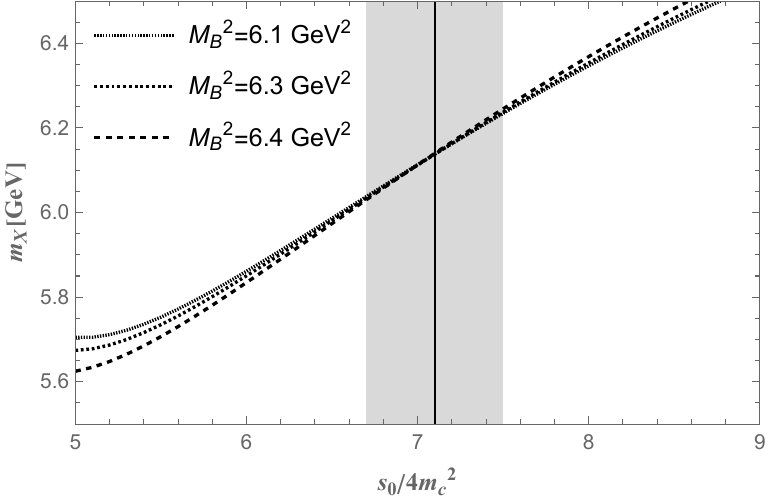}
        \label{fig: mass_J_mu1/2_a}
        }\qquad
    \subfigure[]{
        \includegraphics[width=0.45\textwidth]{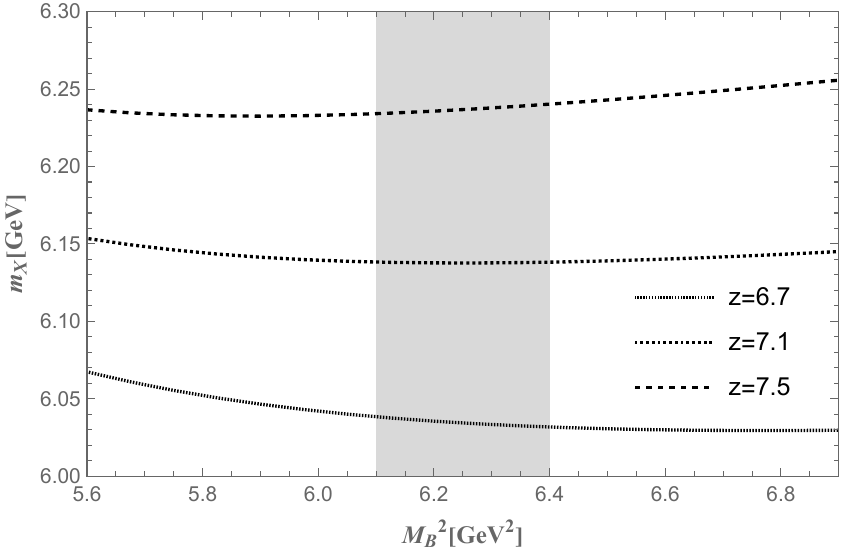}
        \label{fig: mass_J_mu1/2_b}
        }\qquad
    \caption{The variations of charmonium hybrid mass $m_X$ with respect to $z=s_0/4m_c^2$ and $M_B^2$ for $J_{\mu}^{1}$ with $J^{PC}=1^{-+}$.}
    \label{fig: mass_J_mu1/2}
\end{figure}

For the charmonium hybrid system with current $J_{\mu}^{1}$ carrying $J^{PC}=1^{-+}$, we obtain the parameter working regions $6.12$ GeV$^2\leq M_B^2\leq 6.42$ GeV$^2$ and $z=s_0/4m_c^2=7.1\pm 0.4$ after studying the above criteria for the OPE convergence, pole contribution and mass sum rule stability. In figure~\ref{fig: mass_J_mu1/2}, we show the variations of the charmonium hybrid mass $m_X$ with respect to the variables $z=s_{0}/4m_{c}^2$ and $M_{B}^2$. It shows that mass sum rules are very stable within the Borel window (the gray region) in figure~\ref{fig: mass_J_mu1/2_b}, from which one is able to predict the hybrid mass as 
\begin{equation}
m_X=6.14\pm0.19 \text{GeV}\, .
\end{equation}
% The variations for the operator $J_{\mu}^{4/5/6}$ are shown in Fig.~\ref{fig: mass_J_mu4/5/6}, and the mass value is around $7.2$ GeV. By comparison, we can see that even though $J_{\mu}^{1/2}$ and $J_{\mu}^{4/5/6}$ share the same quantum numbers $J^{PC}$, the masses of the states to which they couple are different due to their distinct color structures and distinct contributions from Fig.~\ref{fig: feynman_diagram_g}.
%
%\begin{figure}[htbp]
%    \centering
%    \subfigure[]{
%        \includegraphics[width=0.45\textwidth]{charm(m_vs_z)_J_456mu(1-+).pdf}
%        \label{fig: mass_J_mu4/5/6_a}
%        }\qquad
%    \subfigure[]{
%        \includegraphics[width=0.45\textwidth]{charm(m_vs_MB)_J_456mu(1-+).pdf}
%        \label{fig: mass_J_mu4/5/6_b}
%        }\qquad
%    \caption{The variations of $m_x$ with $z=s_0/4m_c^2$ and $M_B^2$ for the $J_{\mu}^{4/5/6}$ charmonium hybrid.}
%    \label{fig: mass_J_mu4/5/6}
%\end{figure}

After performing the QCD sum rule analyses for all interpolating currents, we calculate the masses of the heavy quarkonium double-gluon hybrid states with $J^{PC}=1^{-+}$ and $2^{+-}$. We summarize the numerical results for the charmonium and bottomonium hybrids in table \ref{table: mass_charmonium} and \ref{table: mass_bottomonium}, respectively. The uncertainties from the heavy quark masses, the strong coupling constants and the condensates are taken into account. 
It shows that the pole contributions for the interpolating currents  $J_{\mu\nu\rho}^1$ and $J_{\mu\nu\rho}^3$ with $J^{PC}=2^{+-}$ are much smaller than those for other currents, implying that these two currents may couple weakly to the heavy quarkonium double-gluon hybrid states. We still give the mass predictions for these two currents, which are much lower than masses extracted from $J_{\mu\nu\rho}^2$ and $J_{\mu\nu\rho}^4$ with the same quantum numbers. 

%There seems to be no discernible order among the masses listed in table \ref{table: mass_charmonium} and \ref{table: mass_bottomonium}. Nevertheless, it seems that some patterns emerge when we classify the color structures of the compact 2g2q states as color symmetry and color antisymmetry. In the case of the charmonium hybrid presented in table \ref{table: mass_charmonium}, we note that the masses of the color symmetric 2g2q states are heavier than those of the color antisymmetric 2g2q states for the same quantum numbers $J^{PC}=1^{-+}$ and $2^{+-}$, respectively, while the situation is inverted for the bottomonium case as presented in table \ref{table: mass_bottomonium}. When considering the color antisymmetric operators, the mass of the charmonium hybrid state with $2^{+-}$ is heavier than that of the charmonium hybrid state with $1^{-+}$, while the situation is also reversed for the bottomonium hybrid case. However, for the color symmetric operators, the mass of the charmonium hybrid state with $2^{+-}$ is lower than that of the charmonium hybrid state with $1^{-+}$, and this situation also appears in the bottomonium scenario.
\begin{table}[htbp] 
    \centering
    \renewcommand\arraystretch{1.8}
    \setlength{\tabcolsep}{2mm}
    {
        \begin{tabular}{ccccccc}
            \hline\hline
        Currents  & $J^{PC}$ & Glueball Operator & $s_{0}/4m_c^2$ & $M_B^2~(\text{GeV}^2)$ & $m_{X}~(\text{GeV})$ & PC\\ \hline
            $J_{\mu}^{1}$ & $1^{-+}$ & A & $7.1\pm 0.4$ & $6.12 - 6.42$ & $6.14\pm 0.19$ & $45.9\%$\\
            $J_{\mu}^{5}$ & $1^{-+}$ & S & $9.7\pm 0.52$ & $8.15 - 8.56$ & $7.21\pm 0.14$ & $46.8\%$\\
            $J_{\mu\nu\rho}^{1}$ & $2^{+-}$ & S & $3.24\pm 0.53$ & $7.75 - 8.75$ & $4.22\pm 0.18$ & $2.2\%$\\
            $J_{\mu\nu\rho}^{2}$ & $2^{+-}$ & S & $7.71\pm 0.44$ & $6.77 - 7.25$ & $6.41\pm 0.17$ & $46.6\%$\\
            $J_{\mu\nu\rho}^{3}$ & $2^{+-}$ & A & $3.74\pm 0.28$ & $5.0 - 5.5$ & $4.55\pm 0.17$ & $12\%$\\
            $J_{\mu\nu\rho}^{4}$ & $2^{+-}$ & A & $7.4\pm 0.37$ & $6.16 - 6.51$ & $6.33\pm 0.17$ & $44.3\%$\\
            \hline\hline
        \end{tabular}
    }
    \caption{Numerical results for the $\bar cGGc$ charmonium double-gluon hybrid states. In the third column, ``A'' and ``S'' represent the antisymmetric and symmetric glueball operators $f^{rst}G^{s}_{\mu\nu}G^{t}_{\alpha\beta}$ and $d^{rst}G^{s}_{\mu\nu}G^{t}_{\alpha\beta}$ in the interpolating currents, respectively. }
\label{table: mass_charmonium}
\end{table}
\begin{table}[htbp] 
    \centering
    \renewcommand\arraystretch{1.8}
    \setlength{\tabcolsep}{2mm}
    {
        \begin{tabular}{ccccccc}
            \hline\hline
        Currents  & $J^{PC}$ & Glueball Operator & $s_{0}/4m_b^2$ & $M_B^2~(\text{GeV}^2)$ & $m_{X}~(\text{GeV})$ & PC\\ \hline
            $J_{\mu}^{1}$ & $1^{-+}$ & A & $3.38\pm 0.09$ & $23.55 - 24.43$ & $14.26\pm 0.19$ & $45.8\%$\\
            $J_{\mu}^{5}$ & $1^{-+}$ & S & $3.04\pm 0.12$ & $18.13 - 19.0$ & $13.71\pm 0.2$ & $44.1\%$\\
            $J_{\mu\nu\rho}^{1}$ & $2^{+-}$ & S & $2.16\pm 0.08$ & $15.31 - 15.92$ & $11.67\pm 0.26$ & $19\%$\\
            $J_{\mu\nu\rho}^{2}$ & $2^{+-}$ & S & $2.58\pm 0.11$ & $14.76 - 15.44$ & $12.58\pm 0.16$ & $49.8\%$\\
            $J_{\mu\nu\rho}^{3}$ & $2^{+-}$ & A & $1.52\pm 0.16$ & $12.85 - 14.65$ & $9.85\pm 0.43$ & $5.3\%$\\
            $J_{\mu\nu\rho}^{4}$ & $2^{+-}$ & A & $2.88\pm 0.13$ & $18.2 - 19.1$ & $13.31\pm 0.19$ & $43.3\%$\\
            \hline\hline
        \end{tabular}
    }
\caption{Numerical results for the $\bar bGGb$ bottomonium double-gluon hybrid states.}
\label{table: mass_bottomonium}
\end{table}

%================================================================================
\section{Summary and discussion}\label{sec5}
%================================================================================
In this work, we have revisited the mass spectra of the heavy quarkonium double-gluon hybrid states with $J^{PC}=1^{-+}$ and $2^{+-}$. Considering the $\bar QGGQ$ operators in the octet-octet $(\mathbf{8}_{[\bar{Q}Q]} \otimes \mathbf{8}_{[GG]})$ color structure, we can construct two independent one Lorentz index currents coupling to  $J^{PC}=1^{-+}$. However, only one current ($J_{\mu\nu}^{1}$) with two Lorentz indices coupling to $J^{PC}=2^{+-}$ can be formed. Nevertheless, our calculations show that the two-point correlation function induced by $J_{\mu\nu}^{1}$ with $J^{PC}=2^{+-}$ contains no non-perturbative contribution up to dimension-8 condensate at the leading order of $\alpha_{s}$. To study the $2^{+-}$ hybrid mesons reliably, we have constructed four independent interpolating currents with three Lorentz indices in this channel, as shown in Eq.~\eqref{eq:currentoperator3}.

We calculate the two-point correlation functions and spectral functions up to dimension-8 condensate, considering the Feynman diagrams in figure~\ref{fig: feynman_diagram}. 
For the $\bar QGGQ$ interpolating currents containing the antisymmetric glueball operator $f^{rst}G^{s}_{\mu\nu}G^{t}_{\alpha\beta}$, our calculations show that the diagram in figure~\ref{fig: feynman_diagram_g} will provide very important contributions to the OPE series and sum rule stabilities. However, it gives no contribution to the correlation functions for the currents with symmetric glueball operator $d^{rst}G^{s}_{\mu\nu}G^{t}_{\alpha\beta}$. We perform the QCD sum rule analyses to both the double-gluon charmonium and bottomonium hybrid mesons with $J^{PC}=1^{-+}$ and $2^{+-}$, and collect the numerical results in table \ref{table: mass_charmonium} and \ref{table: mass_bottomonium}, respectively. Neglecting the weakly coupled states extracted from $J_{\mu\nu\rho}^1$ and $J_{\mu\nu\rho}^3$, we find that the charmonium $\bar cGGc$ hybrid mesons with symmetric glueball operators are heavier than those with antisymmetric glueball operators in both the $1^{-+}$ and $2^{+-}$ channels. However, the situation is reversed for the bottomonium $\bar bGGb$ system, in which the hybrid mesons with symmetric glueball operators are lighter than those with antisymmetric glueball operators. 

\begin{figure}[htbp]
    \centering
    \subfigure[]{
        \includegraphics[width=0.45\textwidth]{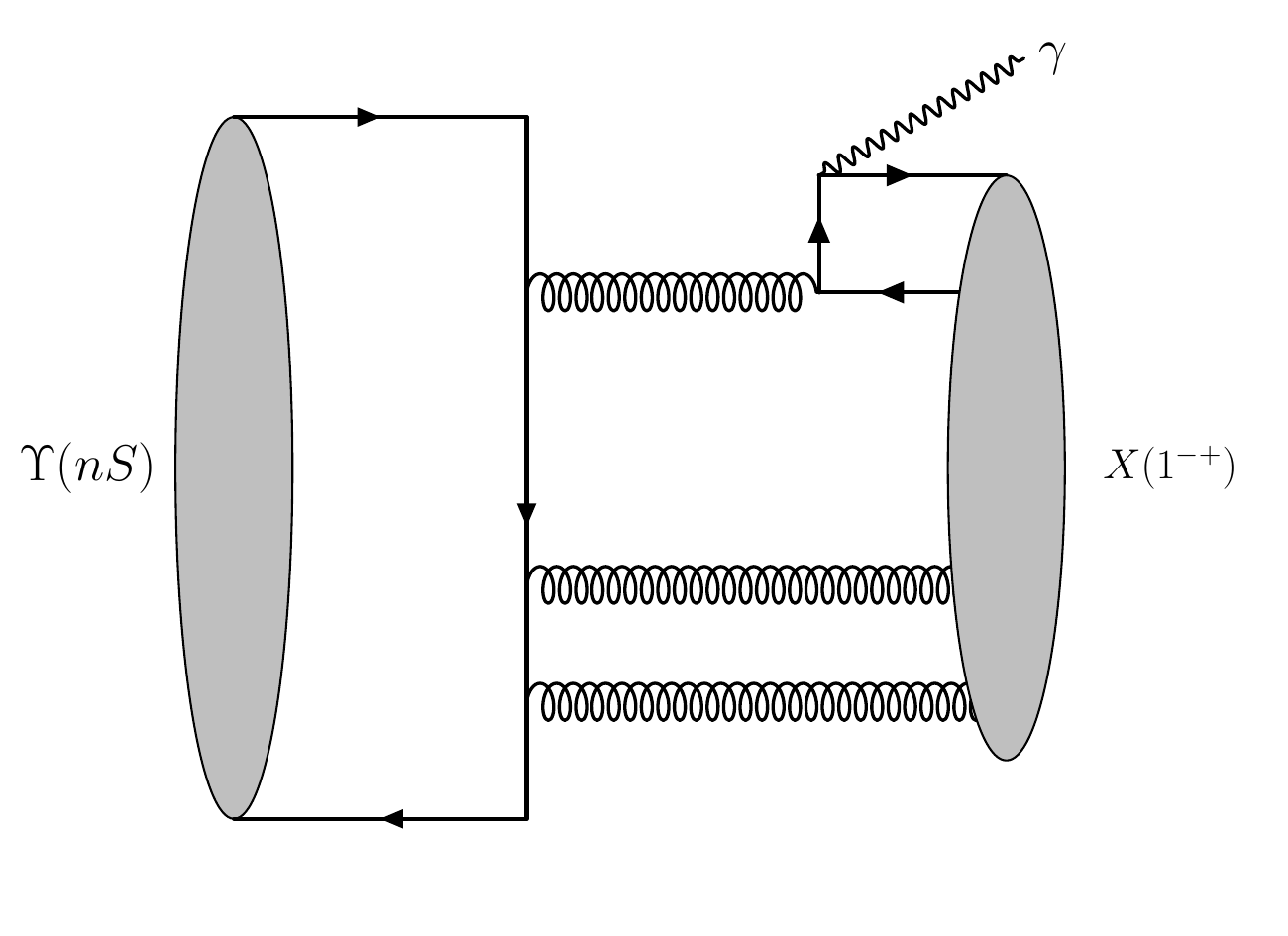}
        \label{fig: production_charmonium_a}
        }\qquad
    \subfigure[]{
        \includegraphics[width=0.45\textwidth]{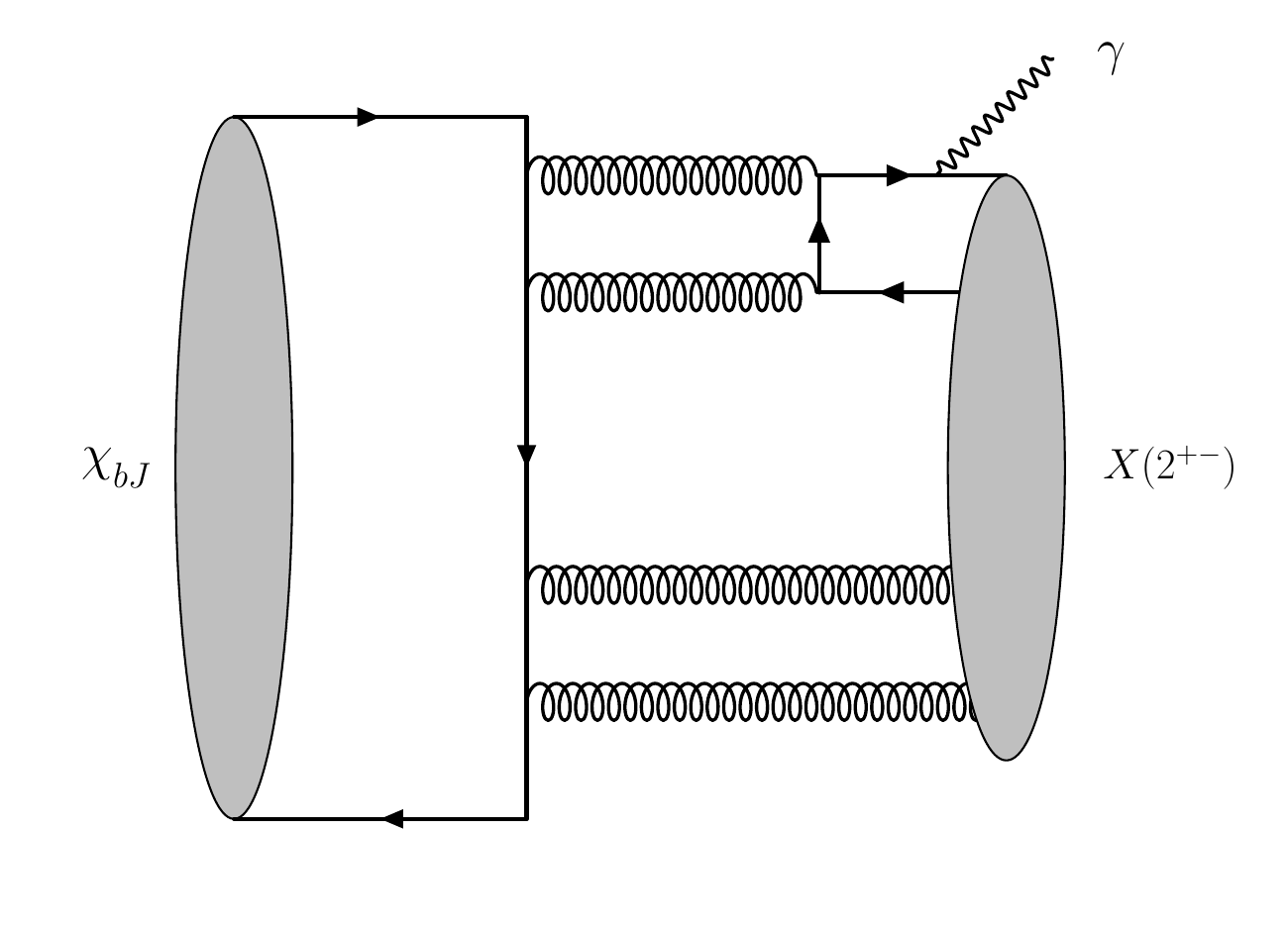}
        \label{fig: production_charmonium_b}
        }\qquad
    \caption{The possible production mechanisms of the charmonium double-gluon hybrids with $J^{PC}=1^{-+}$ and $2^{+-}$.}
    \label{fig: production_charmonium}
\end{figure}

These heavy quarkonium double-gluon hybrid mesons can decay into the final states with two/three mesons or two baryons. One can consult ref.~\cite{Su:2023aif} for detailed discussions. Similar to the production of $\eta_{1}(1855)$ in the $J/\psi$ radiative decay process~\cite{BESIII:2022riz,BESIII:2022iwi,Chen:2022isv}, the charmonium double-gluon hybrids with $J^{PC}=1^{-+}$ and $2^{+-}$ may be produced through the three-gluon and four-gluon emission processes in the hadronic decay of $\Upsilon(n S)$ and $\chi_{bJ}$ respectively, as illustrated in figure~\ref{fig: production_charmonium}. The copious bottomonium mesons in BelleII experiment will be helpful to detect these charmonium double-gluon hybrids in the future. Further investigations on these hybrid mesons in various theoretical and phenomenological methods are also useful and expected.

%================================================================================
\begin{acknowledgments}
This work is supported by the National Natural Science Foundation of China under Grant No. 12175318 and No. 12305147,  the Natural Science Foundation of Guangdong Province of China under Grant No. 2022A1515011922, the National Key R$\&$D Program of China under Contracts No. 2020YFA0406400.
\end{acknowledgments}
%================================================================================

\appendix
%===========================================================
%================================================================================
\section{Operator renormalization}\label{appendix: renormalization}
%================================================================================
The renormalization is performed for the interpolating currents $J_{\mu}^{1}$, $J_{\mu\nu\rho}^{3}$ and $J_{\mu\nu\rho}^{4}$ with the antisymmetric glueball operator $f^{rst}G^{s}_{\mu\nu}G^{t}_{\alpha\beta}$, for which the Feynman diagram in figure~\ref{fig: feynman_diagram_g} will give an important contribution to the tri-gluon condensate. Typically, the diagram in figure~\ref{fig: counter_diagram_a} for the double-gloun hybrid operator has two kinds of vertices. For $J_{\mu}^{1}$, its vertices are $A_{J_{\mu}^{1}}=g_{s}^{2}f^{rst}G^{s}_{\mu\nu}T^{r}\gamma_{\rho}$ and $B_{J_{\mu}^{1}}=g_{s}^{2}f^{rst}G^{t,\,\nu\rho}T^{r}\gamma_{\rho}$. For $J_{\mu\nu\rho}^{3}$, its vertices are 
$A_{J_{\mu\nu\rho}^{3}}=g_{s}^{2}f^{rst}G^{s}_{\mu\nu}T^{r}\gamma_{\alpha}\gamma_{5}$ and $B_{J_{\mu\nu\rho}^{3}}=g_{s}^{2}f^{rst}G^{t,\,\alpha}_{\quad\rho}T^{r}\gamma_{\alpha}\gamma_{5}$. For $J_{\mu\nu\rho}^{4}$, its vertices are $A_{J_{\mu\nu\rho}^{4}}=g_{s}^{2}f^{rst}G^{s}_{\mu\nu}T^{r}\gamma_{\alpha}\gamma_{5}$ and $B_{J_{\mu\nu\rho}^{4}}=g_{s}^{2}f^{rst}\widetilde{G}^{t,\,\alpha}_{\quad\rho}T^{r}\gamma_{\alpha}\gamma_{5}$. It is clearly that $A_{J_{\mu\nu\rho}^{3}}=A_{J_{\mu\nu\rho}^{4}}$ and $\widetilde{B}_{J_{\mu\nu\rho}^{3}}=B_{J_{\mu\nu\rho}^{4}}$, where $\widetilde{B}$ denotes the duality of $B$. We employ the $\overline{\text{MS}}$ scheme to evaluate the renormalization coefficients of the counter terms in figure~\ref{fig: counter_diagram_b} using the dimensional regularization. The results are shown in table~\ref{table: renormalization_constant}.
\begin{table}[htbp]
    \centering
    \setlength{\tabcolsep}{2mm}{
        \renewcommand\arraystretch{1.38}
        \begin{tabular}{ccc}
            \hline\hline
            Vertices of currents & Vertices of counter terms& Renormalization coefficients\\
            \hline
            \multirow{7}{*}{$A_{J_{\mu}^{1}}$} & $i g_s^2 m_Q^2 T^{t}\left(\sigma_{\mu\nu}\gamma_{\rho}-\gamma_{\rho}\sigma_{\mu\nu}\right)$ &-$\frac{3}{4}$\\
            &$g_s^2 m_Q T^{t}\left(k_{1,\,\mu}\gamma_{\rho}\gamma_{\nu}-k_{1,\,\nu}\gamma_{\rho}\gamma_{\mu}\right)$ &$ \frac{3}{4}$\\
            &$i g_s^2 k_1^2 T^{t} \gamma_{\rho}\sigma_{\mu\nu}$ &$-\frac{1}{4}$\\
            &$g_s^2 T^{t}\gamma_{\rho} \left(k_{1,\,\mu}\cancel{k}_{1}\gamma_{\nu}-k_{1,\,\nu}\cancel{k}_{1}\gamma_{\mu}\right)$ & $\frac{1}{4}$\\
            & $g_s^2 m_Q T^{t}\left(k_{2,\,\mu}\gamma_{\nu}\gamma_{\rho}-k_{2,\,\nu}\gamma_{\mu}\gamma_{\rho}\right)$ & $\frac{3}{4}$\\
            &$i g_s^2 k_2^2 T^{t} \sigma_{\mu\nu}\gamma_{\rho}$& $\frac{1}{4}$\\
            &$g_s^2 T^{t}\left(k_{2,\,\mu}\gamma_{\nu}\cancel{k}_{2}-k_{2,\,\nu}\gamma_{\mu}\cancel{k}_{2}\right)\gamma_{\rho}$ &$\frac{1}{4}$\\
            \hline
            \multirow{5}{*}{$B_{J_{\mu}^{1}}$}& $g_s^2 m_Q^2 T^{s} \gamma_{\rho}$ &$-\frac{9}{2}$\\
            &$g_s^2 m_Q T^{s}\left(\cancel{k}_1\gamma_{\rho}+\gamma_{\rho}\cancel{k}_2\right)$ &$\frac{3}{4}$\\
            &$g_s^2 m_Q T^{s}(k_{1,\,\rho}+k_{2,\,\rho})$& $-3$\\
            &$g_s^2 T^{s} \gamma_{\rho}(k_1^2+k_2^2)$ & $1$\\
            &$g_s^2 T^{s}(k_{1,\,\rho}\cancel{k}_{1}+k_{2,\,\rho}\cancel{k}_{2})$ & $\frac{1}{2}$\\
            \hline
            \multirow{7}{*}{$A_{J_{\mu\nu\rho}^{3}}$ (or $A_{J_{\mu\nu\rho}^{4}}$)} & $i g_s^2m_Q^2T^{t}\left(\sigma_{\mu\nu}\gamma_{\alpha}\gamma_5-\gamma_{\alpha}\gamma_5\sigma_{\mu\nu}\right)$&$-\frac{3}{4}$ \\
            &$g_s^2m_QT^{t}\left(k_{1,\,\mu}\gamma_{\alpha}\gamma_5\gamma_{\nu}-k_{1,\,\nu}\gamma_{\alpha}\gamma_5\gamma_{\mu}\right)$ & $\frac{3}{4}$ \\
            &$i g_s^2k_1^2T^{t}\gamma_{\alpha}\gamma_5\sigma_{\mu\nu}$ &$-\frac{1}{4}$\\
            & $g_s^2T^{t}\gamma_{\alpha}\gamma_5\left(k_{1,\,\mu}\cancel{k}_{1}\gamma_{\nu}-k_{1,\,\nu}\cancel{k}_{1}\gamma_{\mu}\right) $& $\frac{1}{4}$\\
            &$g_s^2m_QT^{t}\left(k_{2,\,\mu}\gamma_{\nu}\gamma_{\alpha}\gamma_5-k_{2,\,\nu}\gamma_{\mu}\gamma_{\alpha}\gamma_5\right)$& $\frac{3}{4}$\\
            &$i g_s^2k_2^2T^{t}\sigma_{\mu\nu}\gamma_{\alpha}\gamma_5$ & $\frac{1}{4}$\\
            &$g_s^2T^{t}\left(k_{2,\,\mu}\cancel{k}_{2}\gamma_{\nu}-k_{2,\,\nu}\cancel{k}_{2}\gamma_{\mu}\right)\gamma_{\alpha}\gamma_5$&$\frac{1}{4}$\\
            \hline
            \multirow{5}{*}{$B_{J_{\mu\nu\rho}^{3}}$} & $g_s^{2}m_Q^2 T^{s} \gamma_{\rho}\gamma_5$ & $\frac{9}{2}$\\
            & $ g_s^{2}m_Q T^{s}\left(\cancel{k}_{1}\gamma_{\rho}\gamma_5+\gamma_{\rho}\gamma_5\cancel{k}_{2}\right)$ & $\frac{3}{4}$\\
            & $g_s^{2}m_Q T^{s}\gamma_5\left(k_{1,\,\rho}-k_{2,\rho}\right)$ & $-3$\\
            & $g_s^{2} T^{s} \left(k_{1}^2+k_{2}^2\right)\gamma_{\rho}\gamma_5$ &$-1$\\
            &$g_s^{2} T^{s}\left(k_{1,\,\rho}\cancel{k}_{1}+k_{2,\,\rho}\cancel{k}_{2}\right)\gamma_5$&$-\frac{1}{2}$\\
            \hline
            \multirow{3}{*}{$B_{J_{\mu\nu\rho}^{4}}$} & $g_s^{2}m_Q T^{s} \left(\sigma_{\rho\alpha}k_{1}^{\alpha}-\sigma_{\rho\alpha}k_{2}^{\alpha}\right)$ & $-\frac{3}{2}$\\
            & $i g_s^{2}T^{s}\left(k_{1}^2-k_{2}^2\right)\gamma_{\rho}$ & $\frac{5}{4}$ \\
            & $i g_s^{2}T^{s}\left(k_{1,\,\rho}\cancel{k}_{1}-k_{2,\rho}\cancel{k}_{2}\right)$ &$-\frac{1}{2}$\\
            \hline\hline
        \end{tabular}
    }
    \caption{The vertices and renormalization coefficients of counter terms for the diagram figure~\ref{fig: counter_diagram_b}. $k_1$ and $k_2$ represent the  momenta of incoming and outgoing fermion lines, respectively. An overall factor $\frac{g_s}{16\pi^2}\left(\frac{1}{\epsilon}-\gamma_{E}+\log(4\pi)\right)$ for the renormalization coefficients is understood in the table. }
    \label{table: renormalization_constant}
\end{table}

%================================================================================
\section{Correlation functions and spectral densities}\label{appendix: correlation_spectral}
%================================================================================
In this appendix, we list the correlation functions and spectral densities for the interpolating currents
constructed in section~\ref{sec2}. We omit the polynomials of $z=\frac{q^2}{4m_Q^2}$ in the correlation functions since they will be eliminated by the Borel transform. 
\begin{itemize}
    \item For the interpolating currents with $J^{PC}=1^{-+}$
        \begin{itemize}
            \item[$\bullet $] $J_{\mu}^{1}=-J_{\mu}^{2}$:
                \begin{align*}
                    \begin{autobreak}
                        \Pi^{pert}_{J_{\mu}^{1/2}}=
                        \frac{\alpha _s^2 m_Q^{10} z^6 \, \pFq{3}{2}{1,1,3}{\frac{7}{2},10}{z}}{28350 \pi ^4}
                    +\frac{\alpha _s^2 m_Q^{10} z^6 \, \pFq{4}{3}{1,1,2,4}{\frac{5}{2},5,9}{z}}{315 \pi ^4}
                    -\frac{\alpha _s^2 m_Q^{10} z^6 \, \pFq{4}{3}{1,1,2,4}{\frac{5}{2},5,10}{z}}{567 \pi ^4}
                    -\frac{\alpha _s^2 m_Q^{10} z^6 \, \pFq{4}{3}{1,1,2,4}{\frac{5}{2},6,8}{z}}{1575 \pi ^4}
                    +\frac{\alpha _s^2 m_Q^{10} z^6 \, \pFq{4}{3}{1,1,2,4}{\frac{5}{2},6,9}{z}}{630 \pi ^4}
                    -\frac{2 \alpha _s^2 m_Q^{10} z^6 \, \pFq{4}{3}{1,1,2,4}{\frac{5}{2},6,10}{z}}{2835 \pi ^4}
                    -\frac{4 \alpha _s^2 m_Q^{10} z^6 \, \pFq{4}{3}{1,1,2,4}{3,\frac{7}{2},10}{z}}{4725 \pi ^4}
                    +\frac{\alpha _s^2 m_Q^{10} z^6 \, \pFq{4}{3}{1,1,2,5}{\frac{7}{2},4,10}{z}}{14175 \pi ^4}
                    -\frac{4 \alpha _s^2 m_Q^{10} z^6 \, \pFq{4}{3}{1,1,2,5}{\frac{7}{2},6,10}{z}}{23625 \pi ^4}
                    -\frac{\alpha _s^2 m_Q^{10} z^6 \, \pFq{4}{3}{1,1,3,5}{\frac{7}{2},6,9}{z}}{7875 \pi ^4}
                    +\frac{8 \alpha _s^2 m_Q^{10} z^6 \, \pFq{4}{3}{1,1,3,5}{\frac{7}{2},6,10}{z}}{23625 \pi ^4}
                    -\frac{\alpha _s^2 m_Q^{10} z^6 \, \pFq{4}{3}{1,1,3,5}{\frac{7}{2},7,9}{z}}{47250 \pi ^4}
                    +\frac{2 \alpha _s^2 m_Q^{10} z^6 \, \pFq{4}{3}{1,1,3,5}{\frac{7}{2},7,10}{z}}{42525 \pi ^4}
                    -\frac{\alpha _s^2 m_Q^{10} z^6 \, \pFq{4}{3}{1,1,3,6}{\frac{7}{2},7,10}{z}}{85050 \pi ^4}
                    -\frac{\alpha _s^2 m_Q^{10} z^5 \, \pFq{4}{3}{1,1,1,4}{\frac{5}{2},5,9}{z}}{420 \pi ^4}
                    +\frac{4 \alpha _s^2 m_Q^{10} z^5 \, \pFq{4}{3}{1,1,2,3}{\frac{5}{2},5,8}{z}}{2835 \pi ^4}
                    -\frac{\alpha _s^2 m_Q^{10} z^5 \, \pFq{4}{3}{1,1,2,3}{\frac{5}{2},6,7}{z}}{4050 \pi ^4}
                     -\frac{\alpha _s^2 m_Q^{10} z^5 \, \pFq{4}{3}{1,1,2,4}{\frac{5}{2},6,8}{z}}{9450 \pi ^4}\,,
                    \end{autobreak}\\
                    \Pi^{\GGb}_{J_{\mu}^{1/2}}=& 0\,,\\
                    \Pi^{\GGGb}_{J_{\mu}^{1/2}}=&\frac{\alpha_s\GGGb m_Q^4\left(9(z-1)(420z-109)\,\pFq{3}{2}{1,1,1}{\frac{3}{2},3}{z}+z(8z(23-15z)+109)\,\pFq{3}{2}{1,1,2}{\frac{5}{2},4}{z}\right)}{124416\pi^3}+\\
                    &\frac{\alpha_s\GGGb m_Q^4 z \left(3(5z+4)\log \left(\frac{m_Q^2}{\mu^2}\right)-2(8z+1)\right)\,\pFq{2}{1}{1,1}{\frac{5}{2}}{z}}{648\pi^3}\,,\\
                    \Pi^{R,\,\text{fig.2(g)}}_{J_{\mu}^{1/2}}=& \frac{5 \alpha_s\GGGb m_Q^4 \left(z \left(-24 z^2-16 z+5\right)\, \pFq{3}{2}{1,1,2}{\frac{5}{2},4}{z}+9 \left(12 z^2-17 z+5\right) \,
                    \pFq{3}{2}{1,1,1}{\frac{3}{2},3}{z}\right)}{31104 \pi ^3}+\\
                    & \frac{ \alpha_s\GGGb m_Q^4 z \left(3 (5 z+4) \log \left(\frac{m_Q^2}{\mu
                    ^2}\right)-2 (8 z+1)\right)\,\pFq{2}{1}{1,1}{\frac{5}{2}}{z} }{648 \pi ^3}\,\\
                    \Pi^{\GGb^2}_{J_{\mu}^{1/2}}=&\frac{\GGb^2}{108}m_Q^2 z(2z+1)\,\pFq{2}{1}{1,1}{\frac{5}{2}}{z}\,,\\
                    \begin{autobreak}
                        \rho^{pert}_{J_{\mu}^{1/2}}=
                        \frac{12 \alpha _s^2 m_Q^{10} z^6 \MeijerG{2,0}{2,2}{\frac{7}{2},10}{1,3}{\frac{1}{z}}}{\pi ^{7/2}}
                        +\frac{384 \alpha _s^2 m_Q^{10} z^6 \MeijerG{3,0}{3,3}{\frac{5}{2},5,9}{1,2,4}{\frac{1}{z}}}{\pi ^{7/2}}
                        -\frac{1920 \alpha _s^2 m_Q^{10} z^6 \MeijerG{3,0}{3,3}{\frac{5}{2},5,10}{1,2,4}{\frac{1}{z}}}{\pi ^{7/2}}
                        -\frac{48 \alpha _s^2 m_Q^{10} z^6 \MeijerG{3,0}{3,3}{\frac{5}{2},6,8}{1,2,4}{\frac{1}{z}}}{\pi ^{7/2}}
                        +\frac{960 \alpha _s^2 m_Q^{10} z^6 \MeijerG{3,0}{3,3}{\frac{5}{2},6,9}{1,2,4}{\frac{1}{z}}}{\pi ^{7/2}}
                        -\frac{3840 \alpha _s^2 m_Q^{10} z^6 \MeijerG{3,0}{3,3}{\frac{5}{2},6,10}{1,2,4}{\frac{1}{z}}}{\pi ^{7/2}}
                        -\frac{192 \alpha _s^2 m_Q^{10} z^6 \MeijerG{3,0}{3,3}{3,\frac{7}{2},10}{1,2,4}{\frac{1}{z}}}{\pi ^{7/2}}
                        +\frac{12 \alpha _s^2 m_Q^{10} z^6 \MeijerG{3,0}{3,3}{\frac{7}{2},4,10}{1,2,5}{\frac{1}{z}}}{\pi ^{7/2}}
                        -\frac{24 \alpha _s^2 m_Q^{10} z^6 \MeijerG{3,0}{3,3}{\frac{7}{2},6,9}{1,3,5}{\frac{1}{z}}}{\pi ^{7/2}}
                        -\frac{576 \alpha _s^2 m_Q^{10} z^6 \MeijerG{3,0}{3,3}{\frac{7}{2},6,10}{1,2,5}{\frac{1}{z}}}{\pi ^{7/2}}
                        +\frac{576 \alpha _s^2 m_Q^{10} z^6 \MeijerG{3,0}{3,3}{\frac{7}{2},6,10}{1,3,5}{\frac{1}{z}}}{\pi ^{7/2}}
                        -\frac{24 \alpha _s^2 m_Q^{10} z^6 \MeijerG{3,0}{3,3}{\frac{7}{2},7,9}{1,3,5}{\frac{1}{z}}}{\pi ^{7/2}}
                        +\frac{480 \alpha _s^2 m_Q^{10} z^6 \MeijerG{3,0}{3,3}{\frac{7}{2},7,10}{1,3,5}{\frac{1}{z}}}{\pi ^{7/2}}
                        -\frac{24 \alpha _s^2 m_Q^{10} z^6 \MeijerG{3,0}{3,3}{\frac{7}{2},7,10}{1,3,6}{\frac{1}{z}}}{\pi ^{7/2}}
                        +\frac{64 \alpha _s^2 m_Q^{10} z^5 \MeijerG{3,0}{3,3}{\frac{5}{2},5,8}{1,2,3}{\frac{1}{z}}}{\pi ^{7/2}}
                        -\frac{288 \alpha _s^2 m_Q^{10} z^5 \MeijerG{3,0}{3,3}{\frac{5}{2},5,9}{1,1,4}{\frac{1}{z}}}{\pi ^{7/2}}
                        -\frac{8 \alpha _s^2 m_Q^{10} z^5 \MeijerG{3,0}{3,3}{\frac{5}{2},6,7}{1,2,3}{\frac{1}{z}}}{\pi ^{7/2}}
                        -\frac{8 \alpha _s^2 m_Q^{10} z^5 \MeijerG{3,0}{3,3}{\frac{5}{2},6,8}{1,2,4}{\frac{1}{z}}}{\pi ^{7/2}}\,,
                    \end{autobreak}\\
                    \rho^{\GGb}_{J_{\mu}^{1/2}}=&0\,,\\
                    \rho^{\GGGb}_{J_{\mu}^{1/2}}=&\frac{\alpha_s\GGGb m_Q^4\left(2(z-1)(420z-109)\MeijerG{2,0}{2,2}{\frac{3}{2},3}{1,1}{\frac{1}{z}}+z(8z(23-15z)+109) \MeijerG{2,0}{2,2}{\frac{5}{2},4}{1,2}{\frac{1}{z}}\right)}{27648\pi^{5/2}}+\\
                    &\frac{\alpha_s\GGGb m_Q^4 z\left(3(5z+4)\log \left(\frac{m_Q^2}{\mu^2}\right)-2(8z+1)\right)\MeijerG{1,0}{1,1}{\frac{5}{2}}{1}{\frac{1}{z}}}{864\pi^{5/2}}\,,\\
                    \rho^{R,\,\text{fig.2(g)}}_{J_{\mu}^{1/2}}=& \frac{5 \alpha_s\GGGb m_Q^4 \left(z \left(-24 z^2-16 z+5\right)\, \MeijerG{2,0}{2,2}{\frac{5}{2},4}{1,2}{\frac{1}{z}}+ 2 \left(12 z^2-17 z+5\right) \,
                    \MeijerG{2,0}{2,2}{\frac{3}{2},3}{1,1}{\frac{1}{z}}\right)}{6912\pi^{5/2}}+\\
                    &\frac{\alpha_s\GGGb m_Q^4 z\left(3(5z+4)\log \left(\frac{m_Q^2}{\mu^2}\right)-2(8z+1)\right)\MeijerG{1,0}{1,1}{\frac{5}{2}}{1}{\frac{1}{z}}}{864\pi^{5/2}}\,,\\
                    \rho^{\GGb^2}_{J_{\mu}^{1/2}}=&\frac{\GGb^2 }{144}m_Q^2 z(2z+1)\pi^{\frac{1}{2}}\MeijerG{1,0}{1,1}{\frac{5}{2}}{1}{\frac{1}{z}}\,,
                \end{align*}
            \item[$\bullet $] $-\frac{1}{4}J_{\mu}^{4}=J_{\mu}^{5}=J_{\mu}^{6}$ (a factor $-\frac{1}{4}$ for $J_{\mu}^{4}$ is implicit in the correlation functions and spectral densities):
            \begin{align*}
                \begin{autobreak}
                    \Pi^{pert}_{J_{\mu}^{4/5/6}}=
                    \frac{\alpha _s^2 m_Q^{10} z^6 \, \pFq{4}{3}{1,1,2,4}{\frac{5}{2},6,8}{z}}{945 \pi ^4}
                    -\frac{\alpha _s^2 m_Q^{10} z^6 \, \pFq{4}{3}{1,1,2,4}{\frac{5}{2},6,9}{z}}{1890 \pi ^4}
                    -\frac{\alpha _s^2 m_Q^{10} z^6 \, \pFq{4}{3}{1,1,2,5}{\frac{7}{2},6,9}{z}}{4725 \pi ^4}
                    +\frac{\alpha _s^2 m_Q^{10} z^6 \, \pFq{4}{3}{1,1,3,5}{\frac{7}{2},7,9}{z}}{28350 \pi ^4}
                    -\frac{\alpha _s^2 m_Q^{10} z^5 \, \pFq{4}{3}{1,1,1,4}{\frac{5}{2},5,8}{z}}{378 \pi ^4}
                    -\frac{\alpha _s^2 m_Q^{10} z^5 \, \pFq{4}{3}{1,1,2,3}{\frac{5}{2},6,7}{z}}{810 \pi ^4}\,,
                \end{autobreak}\\
                \Pi^{\GGb}_{J_{\mu}^{4/5/6}}=&0\,,\\
                \Pi^{\GGGb}_{J_{\mu}^{4/5/6}}=&\frac{5\alpha _s \GGGb m_Q^4  \left(\left(-108 z^2+99 z+9\right) \, \pFq{3}{2}{1,1,1}{\frac{3}{2},3}{z}+z \left(-24 z^2+88 z+1\right) \, \pFq{3}{2}{1,1,2}{\frac{5}{2},4}{z}\right)}{41472 \pi ^3}\,,\\
                \Pi^{\GGb^2}_{J_{\mu}^{4/5/6}}=&-\frac{\GGb^2}{648} 5 m_Q^2(z-1) z \,\pFq{2}{1}{1,1}{\frac{5}{2}}{z}\,,\\
                \begin{autobreak}
                    \rho^{pert}_{J_{\mu}^{4/5/6}}=
                    \frac{80 \alpha _s^2 m_Q^{10} z^6 \MeijerG{3,0}{3,3}{\frac{5}{2},6,8}{1,2,4}{\frac{1}{z}}}{\pi ^{7/2}}
                    -\frac{320 \alpha _s^2 m_Q^{10} z^6 \MeijerG{3,0}{3,3}{\frac{5}{2},6,9}{1,2,4}{\frac{1}{z}}}{\pi ^{7/2}}
                    -\frac{80 \alpha _s^2 m_Q^{10} z^6 \MeijerG{3,0}{3,3}{\frac{7}{2},6,9}{1,2,5}{\frac{1}{z}}}{\pi ^{7/2}}
                    +\frac{40 \alpha _s^2 m_Q^{10} z^6 \MeijerG{3,0}{3,3}{\frac{7}{2},7,9}{1,3,5}{\frac{1}{z}}}{\pi ^{7/2}}
                    -\frac{40 \alpha _s^2 m_Q^{10} z^5 \MeijerG{3,0}{3,3}{\frac{5}{2},5,8}{1,1,4}{\frac{1}{z}}}{\pi ^{7/2}}
                    -\frac{40 \alpha _s^2 m_Q^{10} z^5 \MeijerG{3,0}{3,3}{\frac{5}{2},6,7}{1,2,3}{\frac{1}{z}}}{\pi ^{7/2}}\,,
                \end{autobreak}\\
                \rho^{\GGb}_{J_{\mu}^{4/5/6}}=&0\,,\\
                \rho^{\GGGb}_{J_{\mu}^{4/5/6}}=&-\frac{5 \alpha _s \GGGb m_Q^4 \left(\left(24 z^2-22 z-2\right) \MeijerG{2,0}{2,2}{\frac{3}{2},3}{1,1}{\frac{1}{z}}+z \left(24 z^2-88 z-1\right) \MeijerG{2,0}{2,2}{\frac{5}{2},4}{1,2}{\frac{1}{z}}\right)}{9216 \pi ^{5/2}}\,,\\
                \rho^{\GGb^2}_{J_{\mu}^{4/5/6}}=&-\frac{\GGb^2}{864} 5 m_Q^2 z (z-1) \pi^{\frac{1}{2}} \MeijerG{1,0}{1,1}{\frac{5}{2}}{1}{\frac{1}{z}}\,,
            \end{align*}
        \end{itemize} 
    \item For the interpolating currents with $J^{PC}=2^{+-}$ 
        \begin{itemize}
            \item [$\bullet $] $J_{\mu\nu}^{1}$: 
            \begin{align*}
                \begin{autobreak}
                    \Pi^{pert}_{J_{\mu\nu}^{1}}=
                    \frac{\alpha _s^2 m_Q^{10} z^4 \, \pFq{3}{2}{1,1,1}{\frac{5}{2},9}{z}}{12600 \pi ^4}
                    -\frac{\alpha _s^2 m_Q^{10} z^4 \, \pFq{3}{2}{1,1,2}{\frac{5}{2},9}{z}}{63000 \pi ^4}
                    +\frac{16 \alpha _s^2 m_Q^{10} z^6 \, \pFq{3}{2}{1,1,2}{\frac{5}{2},10}{z}}{42525 \pi ^4}
                    -\frac{16 \alpha _s^2 m_Q^{10} z^6 \, \pFq{3}{2}{1,1,2}{\frac{5}{2},11}{z}}{70875 \pi ^4}
                    +\frac{13 \alpha _s^2 m_Q^{10} z^5 \, \pFq{3}{2}{1,1,2}{\frac{7}{2},10}{z}}{354375 \pi ^4}
                    -\frac{76 \alpha _s^2 m_Q^{10} z^6 \, \pFq{3}{2}{1,1,3}{\frac{7}{2},10}{z}}{1063125 \pi ^4}
                    -\frac{\alpha _s^2 m_Q^{10} z^5 \, \pFq{3}{2}{1,1,3}{\frac{7}{2},10}{z}}{50625 \pi ^4}
                    +\frac{422 \alpha _s^2 m_Q^{10} z^6 \, \pFq{3}{2}{1,1,3}{\frac{7}{2},11}{z}}{5315625 \pi ^4}
                    +\frac{44 \alpha _s^2 m_Q^{10} z^6 \, \pFq{3}{2}{1,1,3}{\frac{9}{2},11}{z}}{12403125 \pi ^4}
                    -\frac{\alpha _s^2 m_Q^{10} z^6 \, \pFq{3}{2}{1,1,4}{\frac{9}{2},11}{z}}{118125 \pi ^4}
                    +\frac{2 \alpha _s^2 m_Q^{10} z^6 \, \pFq{3}{2}{1,1,5}{\frac{9}{2},11}{z}}{4134375 \pi ^4}
                    +\frac{2 \alpha _s^2 m_Q^{10} z^6 \, \pFq{3}{2}{1,1,6}{\frac{9}{2},11}{z}}{826875 \pi ^4}
                    -\frac{7 \alpha _s^2 m_Q^{10} z^5 \, \pFq{4}{3}{1,1,1,4}{\frac{5}{2},5,8}{z}}{3375 \pi ^4}
                    +\frac{\alpha _s^2 m_Q^{10} z^4 \, \pFq{4}{3}{1,1,1,4}{\frac{5}{2},5,8}{z}}{15750 \pi ^4}
                    -\frac{23 \alpha _s^2 m_Q^{10} z^5 \, \pFq{4}{3}{1,1,1,4}{\frac{5}{2},5,9}{z}}{47250 \pi ^4}
                    +\frac{2 \alpha _s^2 m_Q^{10} z^5 \, \pFq{4}{3}{1,1,1,4}{\frac{5}{2},5,10}{z}}{14175 \pi ^4}
                    +\frac{8 \alpha _s^2 m_Q^{10} z^5 \, \pFq{4}{3}{1,1,2,3}{\frac{5}{2},4,9}{z}}{14175 \pi ^4}
                    +\frac{88 \alpha _s^2 m_Q^{10} z^5 \, \pFq{4}{3}{1,1,2,3}{\frac{5}{2},5,8}{z}}{70875 \pi ^4}
                    +\frac{38 \alpha _s^2 m_Q^{10} z^5 \, \pFq{4}{3}{1,1,2,3}{\frac{5}{2},6,7}{z}}{50625 \pi ^4}
                    +\frac{22 \alpha _s^2 m_Q^{10} z^6 \, \pFq{4}{3}{1,1,2,4}{\frac{5}{2},5,9}{z}}{23625 \pi ^4}
                    -\frac{2 \alpha _s^2 m_Q^{10} z^5 \, \pFq{4}{3}{1,1,2,4}{\frac{5}{2},5,9}{z}}{7875 \pi ^4}
                    -\frac{2 \alpha _s^2 m_Q^{10} z^6 \, \pFq{4}{3}{1,1,2,4}{\frac{5}{2},5,10}{z}}{6075 \pi ^4}
                    -\frac{8 \alpha _s^2 m_Q^{10} z^6 \, \pFq{4}{3}{1,1,2,4}{\frac{5}{2},5,11}{z}}{70875 \pi ^4}
                    +\frac{76 \alpha _s^2 m_Q^{10} z^6 \, \pFq{4}{3}{1,1,2,4}{\frac{5}{2},6,8}{z}}{118125 \pi ^4}
                    -\frac{11 \alpha _s^2 m_Q^{10} z^5 \, \pFq{4}{3}{1,1,2,4}{\frac{5}{2},6,8}{z}}{118125 \pi ^4}
                    -\frac{16 \alpha _s^2 m_Q^{10} z^6 \, \pFq{4}{3}{1,1,2,4}{\frac{5}{2},6,9}{z}}{118125 \pi ^4}
                    -\frac{22 \alpha _s^2 m_Q^{10} z^6 \, \pFq{4}{3}{1,1,2,4}{\frac{5}{2},6,10}{z}}{212625 \pi ^4}
                    -\frac{136 \alpha _s^2 m_Q^{10} z^6 \, \pFq{4}{3}{1,1,2,4}{3,\frac{7}{2},10}{z}}{354375 \pi ^4}
                    +\frac{8 \alpha _s^2 m_Q^{10} z^6 \, \pFq{4}{3}{1,1,2,4}{3,\frac{7}{2},11}{z}}{354375 \pi ^4}
                    -\frac{32 \alpha _s^2 m_Q^{10} z^6 \, \pFq{4}{3}{1,1,2,5}{3,\frac{7}{2},11}{z}}{354375 \pi ^4}
                    +\frac{4 \alpha _s^2 m_Q^{10} z^6 \, \pFq{4}{3}{1,1,2,5}{\frac{7}{2},4,10}{z}}{151875 \pi ^4}
                    +\frac{4 \alpha _s^2 m_Q^{10} z^6 \, \pFq{4}{3}{1,1,2,5}{\frac{7}{2},4,11}{z}}{590625 \pi ^4}
                    -\frac{14 \alpha _s^2 m_Q^{10} z^6 \, \pFq{4}{3}{1,1,2,5}{\frac{7}{2},6,9}{z}}{84375 \pi ^4}
                    +\frac{\alpha _s^2 m_Q^{10} z^5 \, \pFq{4}{3}{1,1,2,5}{\frac{7}{2},6,9}{z}}{196875 \pi ^4}
                    -\frac{184 \alpha _s^2 m_Q^{10} z^6 \, \pFq{4}{3}{1,1,2,5}{\frac{7}{2},6,10}{z}}{5315625 \pi ^4}
                    +\frac{16 \alpha _s^2 m_Q^{10} z^6 \, \pFq{4}{3}{1,1,2,5}{\frac{7}{2},6,11}{z}}{1771875 \pi ^4}
                    +\frac{16 \alpha _s^2 m_Q^{10} z^6 \, \pFq{4}{3}{1,1,2,6}{\frac{7}{2},4,11}{z}}{354375 \pi ^4}
                    -\frac{2 \alpha _s^2 m_Q^{10} z^6 \, \pFq{4}{3}{1,1,2,7}{4,\frac{9}{2},11}{z}}{826875 \pi ^4}
                    -\frac{22 \alpha _s^2 m_Q^{10} z^6 \, \pFq{4}{3}{1,1,3,5}{\frac{7}{2},6,9}{z}}{590625 \pi ^4}
                    +\frac{\alpha _s^2 m_Q^{10} z^5 \, \pFq{4}{3}{1,1,3,5}{\frac{7}{2},6,9}{z}}{196875 \pi ^4}
                    +\frac{104 \alpha _s^2 m_Q^{10} z^6 \, \pFq{4}{3}{1,1,3,5}{\frac{7}{2},6,10}{z}}{1063125 \pi ^4}
                    +\frac{32 \alpha _s^2 m_Q^{10} z^6 \, \pFq{4}{3}{1,1,3,5}{\frac{7}{2},6,11}{z}}{1771875 \pi ^4}
                    +\frac{76 \alpha _s^2 m_Q^{10} z^6 \, \pFq{4}{3}{1,1,3,5}{\frac{7}{2},7,9}{z}}{1771875 \pi ^4}
                    +\frac{44 \alpha _s^2 m_Q^{10} z^6 \, \pFq{4}{3}{1,1,3,5}{\frac{7}{2},7,10}{z}}{15946875 \pi ^4}
                    +\frac{8 \alpha _s^2 m_Q^{10} z^6 \, \pFq{4}{3}{1,1,3,5}{4,\frac{9}{2},11}{z}}{12403125 \pi ^4}
                    +\frac{8 \alpha _s^2 m_Q^{10} z^6 \, \pFq{4}{3}{1,1,3,6}{\frac{7}{2},5,11}{z}}{354375 \pi ^4}
                    -\frac{22 \alpha _s^2 m_Q^{10} z^6 \, \pFq{4}{3}{1,1,3,6}{\frac{7}{2},7,10}{z}}{3189375 \pi ^4}
                    -\frac{4 \alpha _s^2 m_Q^{10} z^6 \, \pFq{4}{3}{1,1,3,6}{\frac{7}{2},7,11}{z}}{5315625 \pi ^4}
                    -\frac{\alpha _s^2 m_Q^{10} z^6 \, \pFq{4}{3}{1,1,4,7}{\frac{9}{2},6,11}{z}}{1378125 \pi ^4}\,,
                \end{autobreak}\\
                &\Pi^{\GGb}_{J_{\mu\nu}^{1}}=0,\quad \Pi^{\GGGb}_{J_{\mu\nu}^{1}}=0,\quad \Pi^{\GGb^2}_{J_{\mu\nu}^{1}}=0\,,\\
                \begin{autobreak}
                    \rho^{pert}_{J_{\mu\nu}^{1}}=
                     \frac{12 \alpha _s^2 m_Q^{10} z^4 \MeijerG{2,0}{2,2}{\frac{5}{2},9}{1,1}{\frac{1}{z}}}{5 \pi ^{7/2}}
                    -\frac{12 \alpha _s^2 m_Q^{10} z^4 \MeijerG{2,0}{2,2}{\frac{5}{2},9}{1,2}{\frac{1}{z}}}{25 \pi ^{7/2}}
                    +\frac{512 \alpha _s^2 m_Q^{10} z^6 \MeijerG{2,0}{2,2}{\frac{5}{2},10}{1,2}{\frac{1}{z}}}{5 \pi ^{7/2}}
                    -\frac{3072 \alpha _s^2 m_Q^{10} z^6 \MeijerG{2,0}{2,2}{\frac{5}{2},11}{1,2}{\frac{1}{z}}}{5 \pi ^{7/2}}
                    +\frac{624 \alpha _s^2 m_Q^{10} z^5 \MeijerG{2,0}{2,2}{\frac{7}{2},10}{1,2}{\frac{1}{z}}}{25 \pi ^{7/2}}
                    -\frac{608 \alpha _s^2 m_Q^{10} z^6 \MeijerG{2,0}{2,2}{\frac{7}{2},10}{1,3}{\frac{1}{z}}}{25 \pi ^{7/2}}
                    -\frac{168 \alpha _s^2 m_Q^{10} z^5 \MeijerG{2,0}{2,2}{\frac{7}{2},10}{1,3}{\frac{1}{z}}}{25 \pi ^{7/2}}
                    +\frac{6752 \alpha _s^2 m_Q^{10} z^6 \MeijerG{2,0}{2,2}{\frac{7}{2},11}{1,3}{\frac{1}{z}}}{25 \pi ^{7/2}}
                    +\frac{1056 \alpha _s^2 m_Q^{10} z^6 \MeijerG{2,0}{2,2}{\frac{9}{2},11}{1,3}{\frac{1}{z}}}{25 \pi ^{7/2}}
                    -\frac{168 \alpha _s^2 m_Q^{10} z^6 \MeijerG{2,0}{2,2}{\frac{9}{2},11}{1,4}{\frac{1}{z}}}{5 \pi ^{7/2}}
                    +\frac{12 \alpha _s^2 m_Q^{10} z^6 \MeijerG{2,0}{2,2}{\frac{9}{2},11}{1,5}{\frac{1}{z}}}{25 \pi ^{7/2}}
                    +\frac{12 \alpha _s^2 m_Q^{10} z^6 \MeijerG{2,0}{2,2}{\frac{9}{2},11}{1,6}{\frac{1}{z}}}{25 \pi ^{7/2}}
                    +\frac{256 \alpha _s^2 m_Q^{10} z^5 \MeijerG{3,0}{3,3}{\frac{5}{2},4,9}{1,2,3}{\frac{1}{z}}}{5 \pi ^{7/2}}
                    -\frac{784 \alpha _s^2 m_Q^{10} z^5 \MeijerG{3,0}{3,3}{\frac{5}{2},5,8}{1,1,4}{\frac{1}{z}}}{25 \pi ^{7/2}}
                    +\frac{24 \alpha _s^2 m_Q^{10} z^4 \MeijerG{3,0}{3,3}{\frac{5}{2},5,8}{1,1,4}{\frac{1}{z}}}{25 \pi ^{7/2}}
                    +\frac{1408 \alpha _s^2 m_Q^{10} z^5 \MeijerG{3,0}{3,3}{\frac{5}{2},5,8}{1,2,3}{\frac{1}{z}}}{25 \pi ^{7/2}}
                    -\frac{1472 \alpha _s^2 m_Q^{10} z^5 \MeijerG{3,0}{3,3}{\frac{5}{2},5,9}{1,1,4}{\frac{1}{z}}}{25 \pi ^{7/2}}
                    +\frac{2816 \alpha _s^2 m_Q^{10} z^6 \MeijerG{3,0}{3,3}{\frac{5}{2},5,9}{1,2,4}{\frac{1}{z}}}{25 \pi ^{7/2}}
                    -\frac{768 \alpha _s^2 m_Q^{10} z^5 \MeijerG{3,0}{3,3}{\frac{5}{2},5,9}{1,2,4}{\frac{1}{z}}}{25 \pi ^{7/2}}
                    +\frac{768 \alpha _s^2 m_Q^{10} z^5 \MeijerG{3,0}{3,3}{\frac{5}{2},5,10}{1,1,4}{\frac{1}{z}}}{5 \pi ^{7/2}}
                    -\frac{1792 \alpha _s^2 m_Q^{10} z^6 \MeijerG{3,0}{3,3}{\frac{5}{2},5,10}{1,2,4}{\frac{1}{z}}}{5 \pi ^{7/2}}
                    -\frac{6144 \alpha _s^2 m_Q^{10} z^6 \MeijerG{3,0}{3,3}{\frac{5}{2},5,11}{1,2,4}{\frac{1}{z}}}{5 \pi ^{7/2}}
                    +\frac{608 \alpha _s^2 m_Q^{10} z^5 \MeijerG{3,0}{3,3}{\frac{5}{2},6,7}{1,2,3}{\frac{1}{z}}}{25 \pi ^{7/2}}
                    +\frac{1216 \alpha _s^2 m_Q^{10} z^6 \MeijerG{3,0}{3,3}{\frac{5}{2},6,8}{1,2,4}{\frac{1}{z}}}{25 \pi ^{7/2}}
                    -\frac{176 \alpha _s^2 m_Q^{10} z^5 \MeijerG{3,0}{3,3}{\frac{5}{2},6,8}{1,2,4}{\frac{1}{z}}}{25 \pi ^{7/2}}
                    -\frac{2048 \alpha _s^2 m_Q^{10} z^6 \MeijerG{3,0}{3,3}{\frac{5}{2},6,9}{1,2,4}{\frac{1}{z}}}{25 \pi ^{7/2}}
                    -\frac{2816 \alpha _s^2 m_Q^{10} z^6 \MeijerG{3,0}{3,3}{\frac{5}{2},6,10}{1,2,4}{\frac{1}{z}}}{5 \pi ^{7/2}}
                    -\frac{2176 \alpha _s^2 m_Q^{10} z^6 \MeijerG{3,0}{3,3}{3,\frac{7}{2},10}{1,2,4}{\frac{1}{z}}}{25 \pi ^{7/2}}
                    +\frac{256 \alpha _s^2 m_Q^{10} z^6 \MeijerG{3,0}{3,3}{3,\frac{7}{2},11}{1,2,4}{\frac{1}{z}}}{5 \pi ^{7/2}}
                    -\frac{256 \alpha _s^2 m_Q^{10} z^6 \MeijerG{3,0}{3,3}{3,\frac{7}{2},11}{1,2,5}{\frac{1}{z}}}{5 \pi ^{7/2}}
                    +\frac{112 \alpha _s^2 m_Q^{10} z^6 \MeijerG{3,0}{3,3}{\frac{7}{2},4,10}{1,2,5}{\frac{1}{z}}}{25 \pi ^{7/2}}
                    +\frac{288 \alpha _s^2 m_Q^{10} z^6 \MeijerG{3,0}{3,3}{\frac{7}{2},4,11}{1,2,5}{\frac{1}{z}}}{25 \pi ^{7/2}}
                    +\frac{384 \alpha _s^2 m_Q^{10} z^6 \MeijerG{3,0}{3,3}{\frac{7}{2},4,11}{1,2,6}{\frac{1}{z}}}{25 \pi ^{7/2}}
                    +\frac{384 \alpha _s^2 m_Q^{10} z^6 \MeijerG{3,0}{3,3}{\frac{7}{2},5,11}{1,3,6}{\frac{1}{z}}}{25 \pi ^{7/2}}
                    -\frac{1568 \alpha _s^2 m_Q^{10} z^6 \MeijerG{3,0}{3,3}{\frac{7}{2},6,9}{1,2,5}{\frac{1}{z}}}{25 \pi ^{7/2}}
                    +\frac{48 \alpha _s^2 m_Q^{10} z^5 \MeijerG{3,0}{3,3}{\frac{7}{2},6,9}{1,2,5}{\frac{1}{z}}}{25 \pi ^{7/2}}
                    -\frac{176 \alpha _s^2 m_Q^{10} z^6 \MeijerG{3,0}{3,3}{\frac{7}{2},6,9}{1,3,5}{\frac{1}{z}}}{25 \pi ^{7/2}}
                    +\frac{24 \alpha _s^2 m_Q^{10} z^5 \MeijerG{3,0}{3,3}{\frac{7}{2},6,9}{1,3,5}{\frac{1}{z}}}{25 \pi ^{7/2}}
                    -\frac{2944 \alpha _s^2 m_Q^{10} z^6 \MeijerG{3,0}{3,3}{\frac{7}{2},6,10}{1,2,5}{\frac{1}{z}}}{25 \pi ^{7/2}}
                    +\frac{832 \alpha _s^2 m_Q^{10} z^6 \MeijerG{3,0}{3,3}{\frac{7}{2},6,10}{1,3,5}{\frac{1}{z}}}{5 \pi ^{7/2}}
                    +\frac{1536 \alpha _s^2 m_Q^{10} z^6 \MeijerG{3,0}{3,3}{\frac{7}{2},6,11}{1,2,5}{\frac{1}{z}}}{5 \pi ^{7/2}}
                    +\frac{1536 \alpha _s^2 m_Q^{10} z^6 \MeijerG{3,0}{3,3}{\frac{7}{2},6,11}{1,3,5}{\frac{1}{z}}}{5 \pi ^{7/2}}
                    +\frac{1216 \alpha _s^2 m_Q^{10} z^6 \MeijerG{3,0}{3,3}{\frac{7}{2},7,9}{1,3,5}{\frac{1}{z}}}{25 \pi ^{7/2}}
                    +\frac{704 \alpha _s^2 m_Q^{10} z^6 \MeijerG{3,0}{3,3}{\frac{7}{2},7,10}{1,3,5}{\frac{1}{z}}}{25 \pi ^{7/2}}
                    -\frac{352 \alpha _s^2 m_Q^{10} z^6 \MeijerG{3,0}{3,3}{\frac{7}{2},7,10}{1,3,6}{\frac{1}{z}}}{25 \pi ^{7/2}}
                    -\frac{384 \alpha _s^2 m_Q^{10} z^6 \MeijerG{3,0}{3,3}{\frac{7}{2},7,11}{1,3,6}{\frac{1}{z}}}{25 \pi ^{7/2}}
                    -\frac{12 \alpha _s^2 m_Q^{10} z^6 \MeijerG{3,0}{3,3}{4,\frac{9}{2},11}{1,2,7}{\frac{1}{z}}}{25 \pi ^{7/2}}
                    +\frac{48 \alpha _s^2 m_Q^{10} z^6 \MeijerG{3,0}{3,3}{4,\frac{9}{2},11}{1,3,5}{\frac{1}{z}}}{25 \pi ^{7/2}}
                    -\frac{12 \alpha _s^2 m_Q^{10} z^6 \MeijerG{3,0}{3,3}{\frac{9}{2},6,11}{1,4,7}{\frac{1}{z}}}{25 \pi ^{7/2}}\,,
                 \end{autobreak}\\
                 &\rho^{\GGb}_{J_{\mu\nu}^{1}}=0,\quad \rho^{\GGGb}_{J_{\mu\nu}^{1}}=0,\quad \rho^{\GGb^2}_{J_{\mu\nu}^{1}}=0\,,\\
            \end{align*}
        \item[$\bullet$] $J_{\mu\nu\rho}^1$:
            \begin{align*}
                \begin{autobreak}
                    \Pi^{pert}_{J_{\mu\nu\rho}^1}=
                    \frac{37 \alpha _s^2 m_Q^{10} z^4 \, \pFq{3}{2}{1,1,1}{\frac{5}{2},9}{z}}{181440 \pi ^4}
                    -\frac{37 \alpha _s^2 m_Q^{10} z^4 \, \pFq{3}{2}{1,1,2}{\frac{5}{2},9}{z}}{907200 \pi ^4}
                    +\frac{74 \alpha _s^2 m_Q^{10} z^6 \, \pFq{3}{2}{1,1,2}{\frac{5}{2},10}{z}}{76545 \pi ^4}
                    -\frac{74 \alpha _s^2 m_Q^{10} z^6 \, \pFq{3}{2}{1,1,2}{\frac{5}{2},11}{z}}{127575 \pi ^4}
                    +\frac{481 \alpha _s^2 m_Q^{10} z^5 \, \pFq{3}{2}{1,1,2}{\frac{7}{2},10}{z}}{5103000 \pi ^4}
                    -\frac{1079 \alpha _s^2 m_Q^{10} z^6 \, \pFq{3}{2}{1,1,3}{\frac{7}{2},10}{z}}{5103000 \pi ^4}
                    -\frac{37 \alpha _s^2 m_Q^{10} z^5 \, \pFq{3}{2}{1,1,3}{\frac{7}{2},10}{z}}{729000 \pi ^4}
                    +\frac{8233 \alpha _s^2 m_Q^{10} z^6 \, \pFq{3}{2}{1,1,3}{\frac{7}{2},11}{z}}{38272500 \pi ^4}
                    +\frac{407 \alpha _s^2 m_Q^{10} z^6 \, \pFq{3}{2}{1,1,3}{\frac{9}{2},11}{z}}{44651250 \pi ^4}
                    -\frac{149 \alpha _s^2 m_Q^{10} z^6 \, \pFq{3}{2}{1,1,4}{\frac{9}{2},11}{z}}{6615000 \pi ^4}
                    +\frac{37 \alpha _s^2 m_Q^{10} z^6 \, \pFq{3}{2}{1,1,5}{\frac{9}{2},11}{z}}{29767500 \pi ^4}
                    +\frac{37 \alpha _s^2 m_Q^{10} z^6 \, \pFq{3}{2}{1,1,6}{\frac{9}{2},11}{z}}{5953500 \pi ^4}
                    -\frac{13 \alpha _s^2 m_Q^{10} z^5 \, \pFq{4}{3}{1,1,1,4}{\frac{5}{2},5,8}{z}}{340200 \pi ^4}
                    +\frac{\alpha _s^2 m_Q^{10} z^4 \, \pFq{4}{3}{1,1,1,4}{\frac{5}{2},5,8}{z}}{25200 \pi ^4}
                    -\frac{23 \alpha _s^2 m_Q^{10} z^5 \, \pFq{4}{3}{1,1,1,4}{\frac{5}{2},5,9}{z}}{22680 \pi ^4}
                    +\frac{29 \alpha _s^2 m_Q^{10} z^5 \, \pFq{4}{3}{1,1,1,4}{\frac{5}{2},5,10}{z}}{51030 \pi ^4}
                    +\frac{\alpha _s^2 m_Q^{10} z^5 \, \pFq{4}{3}{1,1,2,3}{\frac{5}{2},4,9}{z}}{2835 \pi ^4}
                    -\frac{26 \alpha _s^2 m_Q^{10} z^5 \, \pFq{4}{3}{1,1,2,3}{\frac{5}{2},5,8}{z}}{127575 \pi ^4}
                    -\frac{133 \alpha _s^2 m_Q^{10} z^5 \, \pFq{4}{3}{1,1,2,3}{\frac{5}{2},6,7}{z}}{729000 \pi ^4}
                    -\frac{\alpha _s^2 m_Q^{10} z^6 \, \pFq{4}{3}{1,1,2,4}{\frac{5}{2},5,9}{z}}{9450 \pi ^4}
                    -\frac{\alpha _s^2 m_Q^{10} z^5 \, \pFq{4}{3}{1,1,2,4}{\frac{5}{2},5,9}{z}}{6300 \pi ^4}
                    +\frac{25 \alpha _s^2 m_Q^{10} z^6 \, \pFq{4}{3}{1,1,2,4}{\frac{5}{2},5,10}{z}}{30618 \pi ^4}
                    -\frac{58 \alpha _s^2 m_Q^{10} z^6 \, \pFq{4}{3}{1,1,2,4}{\frac{5}{2},5,11}{z}}{127575 \pi ^4}
                    +\frac{31 \alpha _s^2 m_Q^{10} z^6 \, \pFq{4}{3}{1,1,2,4}{\frac{5}{2},6,8}{z}}{850500 \pi ^4}
                    +\frac{13 \alpha _s^2 m_Q^{10} z^5 \, \pFq{4}{3}{1,1,2,4}{\frac{5}{2},6,8}{z}}{850500 \pi ^4}
                    -\frac{11 \alpha _s^2 m_Q^{10} z^6 \, \pFq{4}{3}{1,1,2,4}{\frac{5}{2},6,9}{z}}{340200 \pi ^4}
                    +\frac{\alpha _s^2 m_Q^{10} z^6 \, \pFq{4}{3}{1,1,2,4}{\frac{5}{2},6,10}{z}}{127575 \pi ^4}
                    -\frac{68 \alpha _s^2 m_Q^{10} z^6 \, \pFq{4}{3}{1,1,2,4}{3,\frac{7}{2},10}{z}}{212625 \pi ^4}
                    -\frac{\alpha _s^2 m_Q^{10} z^6 \, \pFq{4}{3}{1,1,2,4}{3,\frac{7}{2},11}{z}}{127575 \pi ^4}
                    -\frac{148 \alpha _s^2 m_Q^{10} z^6 \, \pFq{4}{3}{1,1,2,5}{3,\frac{7}{2},11}{z}}{637875 \pi ^4}
                    +\frac{31 \alpha _s^2 m_Q^{10} z^6 \, \pFq{4}{3}{1,1,2,5}{\frac{7}{2},4,10}{z}}{2551500 \pi ^4}
                    +\frac{13 \alpha _s^2 m_Q^{10} z^6 \, \pFq{4}{3}{1,1,2,5}{\frac{7}{2},4,11}{z}}{425250 \pi ^4}
                    -\frac{13 \alpha _s^2 m_Q^{10} z^6 \, \pFq{4}{3}{1,1,2,5}{\frac{7}{2},6,9}{z}}{4252500 \pi ^4}
                    +\frac{\alpha _s^2 m_Q^{10} z^5 \, \pFq{4}{3}{1,1,2,5}{\frac{7}{2},6,9}{z}}{315000 \pi ^4}
                    -\frac{46 \alpha _s^2 m_Q^{10} z^6 \, \pFq{4}{3}{1,1,2,5}{\frac{7}{2},6,10}{z}}{637875 \pi ^4}
                    +\frac{116 \alpha _s^2 m_Q^{10} z^6 \, \pFq{4}{3}{1,1,2,5}{\frac{7}{2},6,11}{z}}{3189375 \pi ^4}
                    +\frac{74 \alpha _s^2 m_Q^{10} z^6 \, \pFq{4}{3}{1,1,2,6}{\frac{7}{2},4,11}{z}}{637875 \pi ^4}
                    -\frac{37 \alpha _s^2 m_Q^{10} z^6 \, \pFq{4}{3}{1,1,2,7}{4,\frac{9}{2},11}{z}}{5953500 \pi ^4}
                    +\frac{\alpha _s^2 m_Q^{10} z^6 \, \pFq{4}{3}{1,1,3,5}{\frac{7}{2},6,9}{z}}{236250 \pi ^4}
                    +\frac{\alpha _s^2 m_Q^{10} z^5 \, \pFq{4}{3}{1,1,3,5}{\frac{7}{2},6,9}{z}}{315000 \pi ^4}
                    -\frac{278 \alpha _s^2 m_Q^{10} z^6 \, \pFq{4}{3}{1,1,3,5}{\frac{7}{2},6,10}{z}}{3189375 \pi ^4}
                    +\frac{104 \alpha _s^2 m_Q^{10} z^6 \, \pFq{4}{3}{1,1,3,5}{\frac{7}{2},6,11}{z}}{1063125 \pi ^4}
                    -\frac{121 \alpha _s^2 m_Q^{10} z^6 \, \pFq{4}{3}{1,1,3,5}{\frac{7}{2},7,9}{z}}{25515000 \pi ^4}
                    -\frac{\alpha _s^2 m_Q^{10} z^6 \, \pFq{4}{3}{1,1,3,5}{\frac{7}{2},7,10}{z}}{1366875 \pi ^4}
                    +\frac{2 \alpha _s^2 m_Q^{10} z^6 \, \pFq{4}{3}{1,1,3,5}{4,\frac{9}{2},11}{z}}{3189375 \pi ^4}
                    +\frac{37 \alpha _s^2 m_Q^{10} z^6 \, \pFq{4}{3}{1,1,3,6}{\frac{7}{2},5,11}{z}}{637875 \pi ^4}
                    +\frac{19 \alpha _s^2 m_Q^{10} z^6 \, \pFq{4}{3}{1,1,3,6}{\frac{7}{2},7,10}{z}}{15309000 \pi ^4}
                    -\frac{7 \alpha _s^2 m_Q^{10} z^6 \, \pFq{4}{3}{1,1,3,6}{\frac{7}{2},7,11}{z}}{1366875 \pi ^4}
                    -\frac{37 \alpha _s^2 m_Q^{10} z^6 \, \pFq{4}{3}{1,1,4,7}{\frac{9}{2},6,11}{z}}{19845000 \pi ^4}\,,
                \end{autobreak}\\
                \Pi^{\GGb}_{J_{\mu\nu\rho}^1}=&\frac{\alpha _s \GGb m_Q^6 \left(35-8 z (z (48 z (11 z-72)-755)+183)\right) \, \pFq{3}{2}{1,1,2}{\frac{5}{2},4}{z}}{746496 \pi ^2}\\
                &-\frac{\alpha _s \GGb m_Q^6 (z-1) (4 z (24 z (22 z+51)-359)+35) \, \pFq{3}{2}{1,1,1}{\frac{3}{2},3}{z}}{82944 \pi ^2 z}\,,\\
                \Pi^{\GGGb}_{J_{\mu\nu\rho}^1}=&\frac{\alpha _s \GGGb m_Q^4  \left(44016 z^3-71980 z^2+28879 z-915\right) \, \pFq{3}{2}{1,1,1}{\frac{3}{2},3}{z}}{1327104 \pi ^3 z}\\
                &+\frac{\alpha _s \GGGb m_Q^4 \left(74112 z^3-152168 z^2+28696 z-915\right) \, \pFq{3}{2}{1,1,2}{\frac{5}{2},4}{z}}{11943936 \pi ^3}\,,\\
                \Pi^{\GGb^2}_{J_{\mu\nu\rho}^1}=&-\frac{\GGb^2}{1944} 5 m_Q^2 z (2 z+1)\, \pFq{2}{1}{1,1}{\frac{5}{2}}{z}\,,\\
                \begin{autobreak}
                    \rho^{pert}_{J_{\mu\nu\rho}^1}=
                    \frac{37 \alpha _s^2 m_Q^{10} z^4 \MeijerG{2,0}{2,2}{\frac{5}{2},9}{1,1}{\frac{1}{z}}}{6 \pi ^{7/2}}
                    -\frac{37 \alpha _s^2 m_Q^{10} z^4 \MeijerG{2,0}{2,2}{\frac{5}{2},9}{1,2}{\frac{1}{z}}}{30 \pi ^{7/2}}
                    +\frac{2368 \alpha _s^2 m_Q^{10} z^6 \MeijerG{2,0}{2,2}{\frac{5}{2},10}{1,2}{\frac{1}{z}}}{9 \pi ^{7/2}}
                    -\frac{4736 \alpha _s^2 m_Q^{10} z^6 \MeijerG{2,0}{2,2}{\frac{5}{2},11}{1,2}{\frac{1}{z}}}{3 \pi ^{7/2}}
                    +\frac{962 \alpha _s^2 m_Q^{10} z^5 \MeijerG{2,0}{2,2}{\frac{7}{2},10}{1,2}{\frac{1}{z}}}{15 \pi ^{7/2}}
                    -\frac{1079 \alpha _s^2 m_Q^{10} z^6 \MeijerG{2,0}{2,2}{\frac{7}{2},10}{1,3}{\frac{1}{z}}}{15 \pi ^{7/2}}
                    -\frac{259 \alpha _s^2 m_Q^{10} z^5 \MeijerG{2,0}{2,2}{\frac{7}{2},10}{1,3}{\frac{1}{z}}}{15 \pi ^{7/2}}
                    +\frac{32932 \alpha _s^2 m_Q^{10} z^6 \MeijerG{2,0}{2,2}{\frac{7}{2},11}{1,3}{\frac{1}{z}}}{45 \pi ^{7/2}}
                    +\frac{1628 \alpha _s^2 m_Q^{10} z^6 \MeijerG{2,0}{2,2}{\frac{9}{2},11}{1,3}{\frac{1}{z}}}{15 \pi ^{7/2}}
                    -\frac{447 \alpha _s^2 m_Q^{10} z^6 \MeijerG{2,0}{2,2}{\frac{9}{2},11}{1,4}{\frac{1}{z}}}{5 \pi ^{7/2}}
                    +\frac{37 \alpha _s^2 m_Q^{10} z^6 \MeijerG{2,0}{2,2}{\frac{9}{2},11}{1,5}{\frac{1}{z}}}{30 \pi ^{7/2}}
                    +\frac{37 \alpha _s^2 m_Q^{10} z^6 \MeijerG{2,0}{2,2}{\frac{9}{2},11}{1,6}{\frac{1}{z}}}{30 \pi ^{7/2}}
                    +\frac{32 \alpha _s^2 m_Q^{10} z^5 \MeijerG{3,0}{3,3}{\frac{5}{2},4,9}{1,2,3}{\frac{1}{z}}}{\pi ^{7/2}}
                    -\frac{26 \alpha _s^2 m_Q^{10} z^5 \MeijerG{3,0}{3,3}{\frac{5}{2},5,8}{1,1,4}{\frac{1}{z}}}{45 \pi ^{7/2}}
                    +\frac{3 \alpha _s^2 m_Q^{10} z^4 \MeijerG{3,0}{3,3}{\frac{5}{2},5,8}{1,1,4}{\frac{1}{z}}}{5 \pi ^{7/2}}
                    -\frac{416 \alpha _s^2 m_Q^{10} z^5 \MeijerG{3,0}{3,3}{\frac{5}{2},5,8}{1,2,3}{\frac{1}{z}}}{45 \pi ^{7/2}}
                    -\frac{368 \alpha _s^2 m_Q^{10} z^5 \MeijerG{3,0}{3,3}{\frac{5}{2},5,9}{1,1,4}{\frac{1}{z}}}{3 \pi ^{7/2}}
                    -\frac{64 \alpha _s^2 m_Q^{10} z^6 \MeijerG{3,0}{3,3}{\frac{5}{2},5,9}{1,2,4}{\frac{1}{z}}}{5 \pi ^{7/2}}
                    -\frac{96 \alpha _s^2 m_Q^{10} z^5 \MeijerG{3,0}{3,3}{\frac{5}{2},5,9}{1,2,4}{\frac{1}{z}}}{5 \pi ^{7/2}}
                    +\frac{1856 \alpha _s^2 m_Q^{10} z^5 \MeijerG{3,0}{3,3}{\frac{5}{2},5,10}{1,1,4}{\frac{1}{z}}}{3 \pi ^{7/2}}
                    +\frac{8000 \alpha _s^2 m_Q^{10} z^6 \MeijerG{3,0}{3,3}{\frac{5}{2},5,10}{1,2,4}{\frac{1}{z}}}{9 \pi ^{7/2}}
                    -\frac{14848 \alpha _s^2 m_Q^{10} z^6 \MeijerG{3,0}{3,3}{\frac{5}{2},5,11}{1,2,4}{\frac{1}{z}}}{3 \pi ^{7/2}}
                    -\frac{266 \alpha _s^2 m_Q^{10} z^5 \MeijerG{3,0}{3,3}{\frac{5}{2},6,7}{1,2,3}{\frac{1}{z}}}{45 \pi ^{7/2}}
                    +\frac{124 \alpha _s^2 m_Q^{10} z^6 \MeijerG{3,0}{3,3}{\frac{5}{2},6,8}{1,2,4}{\frac{1}{z}}}{45 \pi ^{7/2}}
                    +\frac{52 \alpha _s^2 m_Q^{10} z^5 \MeijerG{3,0}{3,3}{\frac{5}{2},6,8}{1,2,4}{\frac{1}{z}}}{45 \pi ^{7/2}}
                    -\frac{176 \alpha _s^2 m_Q^{10} z^6 \MeijerG{3,0}{3,3}{\frac{5}{2},6,9}{1,2,4}{\frac{1}{z}}}{9 \pi ^{7/2}}
                    +\frac{128 \alpha _s^2 m_Q^{10} z^6 \MeijerG{3,0}{3,3}{\frac{5}{2},6,10}{1,2,4}{\frac{1}{z}}}{3 \pi ^{7/2}}
                    -\frac{1088 \alpha _s^2 m_Q^{10} z^6 \MeijerG{3,0}{3,3}{3,\frac{7}{2},10}{1,2,4}{\frac{1}{z}}}{15 \pi ^{7/2}}
                    -\frac{160 \alpha _s^2 m_Q^{10} z^6 \MeijerG{3,0}{3,3}{3,\frac{7}{2},11}{1,2,4}{\frac{1}{z}}}{9 \pi ^{7/2}}
                    -\frac{1184 \alpha _s^2 m_Q^{10} z^6 \MeijerG{3,0}{3,3}{3,\frac{7}{2},11}{1,2,5}{\frac{1}{z}}}{9 \pi ^{7/2}}
                    +\frac{31 \alpha _s^2 m_Q^{10} z^6 \MeijerG{3,0}{3,3}{\frac{7}{2},4,10}{1,2,5}{\frac{1}{z}}}{15 \pi ^{7/2}}
                    +\frac{52 \alpha _s^2 m_Q^{10} z^6 \MeijerG{3,0}{3,3}{\frac{7}{2},4,11}{1,2,5}{\frac{1}{z}}}{\pi ^{7/2}}
                    +\frac{592 \alpha _s^2 m_Q^{10} z^6 \MeijerG{3,0}{3,3}{\frac{7}{2},4,11}{1,2,6}{\frac{1}{z}}}{15 \pi ^{7/2}}
                    +\frac{592 \alpha _s^2 m_Q^{10} z^6 \MeijerG{3,0}{3,3}{\frac{7}{2},5,11}{1,3,6}{\frac{1}{z}}}{15 \pi ^{7/2}}
                    -\frac{52 \alpha _s^2 m_Q^{10} z^6 \MeijerG{3,0}{3,3}{\frac{7}{2},6,9}{1,2,5}{\frac{1}{z}}}{45 \pi ^{7/2}}
                    +\frac{6 \alpha _s^2 m_Q^{10} z^5 \MeijerG{3,0}{3,3}{\frac{7}{2},6,9}{1,2,5}{\frac{1}{z}}}{5 \pi ^{7/2}}
                    +\frac{4 \alpha _s^2 m_Q^{10} z^6 \MeijerG{3,0}{3,3}{\frac{7}{2},6,9}{1,3,5}{\frac{1}{z}}}{5 \pi ^{7/2}}
                    +\frac{3 \alpha _s^2 m_Q^{10} z^5 \MeijerG{3,0}{3,3}{\frac{7}{2},6,9}{1,3,5}{\frac{1}{z}}}{5 \pi ^{7/2}}
                    -\frac{736 \alpha _s^2 m_Q^{10} z^6 \MeijerG{3,0}{3,3}{\frac{7}{2},6,10}{1,2,5}{\frac{1}{z}}}{3 \pi ^{7/2}}
                    -\frac{2224 \alpha _s^2 m_Q^{10} z^6 \MeijerG{3,0}{3,3}{\frac{7}{2},6,10}{1,3,5}{\frac{1}{z}}}{15 \pi ^{7/2}}
                    +\frac{3712 \alpha _s^2 m_Q^{10} z^6 \MeijerG{3,0}{3,3}{\frac{7}{2},6,11}{1,2,5}{\frac{1}{z}}}{3 \pi ^{7/2}}
                    +\frac{1664 \alpha _s^2 m_Q^{10} z^6 \MeijerG{3,0}{3,3}{\frac{7}{2},6,11}{1,3,5}{\frac{1}{z}}}{\pi ^{7/2}}
                    -\frac{242 \alpha _s^2 m_Q^{10} z^6 \MeijerG{3,0}{3,3}{\frac{7}{2},7,9}{1,3,5}{\frac{1}{z}}}{45 \pi ^{7/2}}
                    -\frac{112 \alpha _s^2 m_Q^{10} z^6 \MeijerG{3,0}{3,3}{\frac{7}{2},7,10}{1,3,5}{\frac{1}{z}}}{15 \pi ^{7/2}}
                    +\frac{38 \alpha _s^2 m_Q^{10} z^6 \MeijerG{3,0}{3,3}{\frac{7}{2},7,10}{1,3,6}{\frac{1}{z}}}{15 \pi ^{7/2}}
                    -\frac{1568 \alpha _s^2 m_Q^{10} z^6 \MeijerG{3,0}{3,3}{\frac{7}{2},7,11}{1,3,6}{\frac{1}{z}}}{15 \pi ^{7/2}}
                    -\frac{37 \alpha _s^2 m_Q^{10} z^6 \MeijerG{3,0}{3,3}{4,\frac{9}{2},11}{1,2,7}{\frac{1}{z}}}{30 \pi ^{7/2}}
                    +\frac{28 \alpha _s^2 m_Q^{10} z^6 \MeijerG{3,0}{3,3}{4,\frac{9}{2},11}{1,3,5}{\frac{1}{z}}}{15 \pi ^{7/2}}
                    -\frac{37 \alpha _s^2 m_Q^{10} z^6 \MeijerG{3,0}{3,3}{\frac{9}{2},6,11}{1,4,7}{\frac{1}{z}}}{30 \pi ^{7/2}}\,,
                \end{autobreak}\\
                \rho^{\GGb}_{J_{\mu\nu\rho}^1}=&-\frac{\alpha _s \GGb m_Q^6 (4 z (24 z (22 z+29)-1583)+1471) \MeijerG{2,0}{2,2}{\frac{3}{2},3}{1,1}{\frac{1}{z}}}{82944 \pi ^{3/2}}\\
                &+\frac{\alpha _s \GGb m_Q^6 \left((35-8 z (z (48 z (11 z-72)-755)+183)) \MeijerG{2,0}{2,2}{\frac{5}{2},4}{1,2}{\frac{1}{z}}+70 \MeijerG{2,0}{2,2}{\frac{5}{2},4}{2,2}{\frac{1}{z}}\right)}{165888 \pi ^{3/2}}\,,\\
                \rho^{\GGGb}_{J_{\mu\nu\rho}^1}=&\frac{\alpha _s \GGGb m_Q^4 \left(\left(44016 z^2-71980 z+28879\right) \MeijerG{2,0}{2,2}{\frac{3}{2},3}{1,1}{\frac{1}{z}}-915 \MeijerG{2,0}{2,2}{\frac{5}{2},4}{2,2}{\frac{1}{z}}\right)}{1327104 \pi ^{5/2}}\\
                &+\frac{\alpha _s \GGGb m_Q^4 \left(74112 z^3-152168 z^2+28696 z-915\right) \MeijerG{2,0}{2,2}{\frac{5}{2},4}{1,2}{\frac{1}{z}}}{2654208 \pi ^{5/2}}\,,\\
                \rho^{\GGb^2}_{J_{\mu\nu\rho}^1}=&-\frac{\GGb^2 }{2592}5 m_Q^2 z (2 z+1) \pi^{\frac{1}{2}} \MeijerG{1,0}{1,1}{\frac{5}{2}}{1}{\frac{1}{z}}\,,
            \end{align*}
        \item[$\bullet$] $J_{\mu\nu\rho}^2$:
            \begin{align*}
               \Pi^{pert}_{J_{\mu\nu\rho}^2}=& \Pi^{pert}_{J_{\mu\nu\rho}^1},\quad \Pi^{\GGb}_{J_{\mu\nu\rho}^2}=-\Pi^{\GGb}_{J_{\mu\nu\rho}^1},\quad \Pi^{\GGb^2}_{J_{\mu\nu\rho}^2}=\Pi^{\GGb^2}_{J_{\mu\nu\rho}^1}\,,\\
                \Pi^{\GGGb}_{J_{\mu\nu\rho}^2}=&\frac{\alpha _s \GGGb m_Q^4 \left(915-8 z (z (7584 z-22141)+3497)\right) \, \pFq{3}{2}{1,1,2}{\frac{5}{2},4}{z}}{11943936 \pi ^3}\\
                &-\frac{\alpha _s \GGGb m_Q^4 (z-1) (28 z (1332 z-973)+915) \, \pFq{3}{2}{1,1,1}{\frac{3}{2},3}{z}}{1327104 \pi ^3 z}\,,\\
                \rho^{pert}_{J_{\mu\nu\rho}^2}=&\rho^{pert}_{J_{\mu\nu\rho}^1},\quad \rho^{\GGb}_{J_{\mu\nu\rho}^2}=-\rho^{\GGb}_{J_{\mu\nu\rho}^1},\quad \rho^{\GGb^2}_{J_{\mu\nu\rho}^2}=\rho^{\GGb^2}_{J_{\mu\nu\rho}^1}\,,\\
                \rho^{\GGGb}_{J_{\mu\nu\rho}^2}=&\frac{\alpha _s \GGGb m_Q^4 \left(-\left(37296 z^2-64540 z+28159\right) \MeijerG{2,0}{2,2}{\frac{3}{2},3}{1,1}{\frac{1}{z}}+915 \MeijerG{2,0}{2,2}{\frac{5}{2},4}{2,2}{\frac{1}{z}}\right)}{1327104 \pi ^{5/2}}\\
                &+\frac{\alpha _s \GGGb m_Q^4 \left(-60672 z^3+177128 z^2-27976 z+915\right) \MeijerG{2,0}{2,2}{\frac{5}{2},4}{1,2}{\frac{1}{z}}}{2654208 \pi ^{5/2}}\,,
            \end{align*}
        \item[$\bullet$] $J_{\mu\nu\rho}^3$:
            \begin{align*}
                \begin{autobreak}
                    \Pi^{pert}_{J_{\mu\nu\rho}^3}=
                    \frac{\alpha _s^2 m_Q^{10} z^4 \, \pFq{3}{2}{1,1,1}{\frac{5}{2},9}{z}}{14400 \pi ^4}
                    -\frac{\alpha _s^2 m_Q^{10} z^4 \, \pFq{3}{2}{1,1,2}{\frac{5}{2},9}{z}}{72000 \pi ^4}
                    +\frac{2 \alpha _s^2 m_Q^{10} z^6 \, \pFq{3}{2}{1,1,2}{\frac{5}{2},10}{z}}{6075 \pi ^4}
                    -\frac{2 \alpha _s^2 m_Q^{10} z^6 \, \pFq{3}{2}{1,1,2}{\frac{5}{2},11}{z}}{10125 \pi ^4}
                    +\frac{13 \alpha _s^2 m_Q^{10} z^5 \, \pFq{3}{2}{1,1,2}{\frac{7}{2},10}{z}}{405000 \pi ^4}
                    -\frac{557 \alpha _s^2 m_Q^{10} z^6 \, \pFq{3}{2}{1,1,3}{\frac{7}{2},10}{z}}{8505000 \pi ^4}
                    -\frac{7 \alpha _s^2 m_Q^{10} z^5 \, \pFq{3}{2}{1,1,3}{\frac{7}{2},10}{z}}{405000 \pi ^4}
                    +\frac{1507 \alpha _s^2 m_Q^{10} z^6 \, \pFq{3}{2}{1,1,3}{\frac{7}{2},11}{z}}{21262500 \pi ^4}
                    +\frac{11 \alpha _s^2 m_Q^{10} z^6 \, \pFq{3}{2}{1,1,3}{\frac{9}{2},11}{z}}{3543750 \pi ^4}
                    -\frac{17 \alpha _s^2 m_Q^{10} z^6 \, \pFq{3}{2}{1,1,4}{\frac{9}{2},11}{z}}{2205000 \pi ^4}
                    +\frac{\alpha _s^2 m_Q^{10} z^6 \, \pFq{3}{2}{1,1,5}{\frac{9}{2},11}{z}}{2362500 \pi ^4}
                    +\frac{\alpha _s^2 m_Q^{10} z^6 \, \pFq{3}{2}{1,1,6}{\frac{9}{2},11}{z}}{472500 \pi ^4}
                    -\frac{43 \alpha _s^2 m_Q^{10} z^5 \, \pFq{4}{3}{1,1,1,4}{\frac{5}{2},5,8}{z}}{189000 \pi ^4}
                    -\frac{\alpha _s^2 m_Q^{10} z^4 \, \pFq{4}{3}{1,1,1,4}{\frac{5}{2},5,8}{z}}{42000 \pi ^4}
                    -\frac{17 \alpha _s^2 m_Q^{10} z^5 \, \pFq{4}{3}{1,1,1,4}{\frac{5}{2},5,9}{z}}{27000 \pi ^4}
                    +\frac{\alpha _s^2 m_Q^{10} z^5 \, \pFq{4}{3}{1,1,1,4}{\frac{5}{2},5,10}{z}}{5670 \pi ^4}
                    -\frac{\alpha _s^2 m_Q^{10} z^5 \, \pFq{4}{3}{1,1,2,3}{\frac{5}{2},4,9}{z}}{4725 \pi ^4}
                    -\frac{8 \alpha _s^2 m_Q^{10} z^5 \, \pFq{4}{3}{1,1,2,3}{\frac{5}{2},5,8}{z}}{70875 \pi ^4}
                    +\frac{161 \alpha _s^2 m_Q^{10} z^5 \, \pFq{4}{3}{1,1,2,3}{\frac{5}{2},6,7}{z}}{405000 \pi ^4}
                    +\frac{13 \alpha _s^2 m_Q^{10} z^6 \, \pFq{4}{3}{1,1,2,4}{\frac{5}{2},5,9}{z}}{23625 \pi ^4}
                    +\frac{\alpha _s^2 m_Q^{10} z^5 \, \pFq{4}{3}{1,1,2,4}{\frac{5}{2},5,9}{z}}{10500 \pi ^4}
                    -\frac{\alpha _s^2 m_Q^{10} z^6 \, \pFq{4}{3}{1,1,2,4}{\frac{5}{2},5,10}{z}}{14175 \pi ^4}
                    -\frac{2 \alpha _s^2 m_Q^{10} z^6 \, \pFq{4}{3}{1,1,2,4}{\frac{5}{2},5,11}{z}}{14175 \pi ^4}
                    -\frac{\alpha _s^2 m_Q^{10} z^6 \, \pFq{4}{3}{1,1,2,4}{\frac{5}{2},6,8}{z}}{52500 \pi ^4}
                    +\frac{\alpha _s^2 m_Q^{10} z^5 \, \pFq{4}{3}{1,1,2,4}{\frac{5}{2},6,8}{z}}{118125 \pi ^4}
                    +\frac{179 \alpha _s^2 m_Q^{10} z^6 \, \pFq{4}{3}{1,1,2,4}{\frac{5}{2},6,9}{z}}{945000 \pi ^4}
                    -\frac{17 \alpha _s^2 m_Q^{10} z^6 \, \pFq{4}{3}{1,1,2,4}{\frac{5}{2},6,10}{z}}{170100 \pi ^4}
                    -\frac{94 \alpha _s^2 m_Q^{10} z^6 \, \pFq{4}{3}{1,1,2,4}{3,\frac{7}{2},10}{z}}{354375 \pi ^4}
                    +\frac{\alpha _s^2 m_Q^{10} z^6 \, \pFq{4}{3}{1,1,2,4}{3,\frac{7}{2},11}{z}}{354375 \pi ^4}
                    -\frac{4 \alpha _s^2 m_Q^{10} z^6 \, \pFq{4}{3}{1,1,2,5}{3,\frac{7}{2},11}{z}}{50625 \pi ^4}
                    +\frac{73 \alpha _s^2 m_Q^{10} z^6 \, \pFq{4}{3}{1,1,2,5}{\frac{7}{2},4,10}{z}}{4252500 \pi ^4}
                    +\frac{11 \alpha _s^2 m_Q^{10} z^6 \, \pFq{4}{3}{1,1,2,5}{\frac{7}{2},4,11}{z}}{1181250 \pi ^4}
                    -\frac{43 \alpha _s^2 m_Q^{10} z^6 \, \pFq{4}{3}{1,1,2,5}{\frac{7}{2},6,9}{z}}{2362500 \pi ^4}
                    -\frac{\alpha _s^2 m_Q^{10} z^5 \, \pFq{4}{3}{1,1,2,5}{\frac{7}{2},6,9}{z}}{525000 \pi ^4}
                    -\frac{34 \alpha _s^2 m_Q^{10} z^6 \, \pFq{4}{3}{1,1,2,5}{\frac{7}{2},6,10}{z}}{759375 \pi ^4}
                    +\frac{4 \alpha _s^2 m_Q^{10} z^6 \, \pFq{4}{3}{1,1,2,5}{\frac{7}{2},6,11}{z}}{354375 \pi ^4}
                    +\frac{2 \alpha _s^2 m_Q^{10} z^6 \, \pFq{4}{3}{1,1,2,6}{\frac{7}{2},4,11}{z}}{50625 \pi ^4}
                    -\frac{\alpha _s^2 m_Q^{10} z^6 \, \pFq{4}{3}{1,1,2,7}{4,\frac{9}{2},11}{z}}{472500 \pi ^4}
                    -\frac{13 \alpha _s^2 m_Q^{10} z^6 \, \pFq{4}{3}{1,1,3,5}{\frac{7}{2},6,9}{z}}{590625 \pi ^4}
                    -\frac{\alpha _s^2 m_Q^{10} z^5 \, \pFq{4}{3}{1,1,3,5}{\frac{7}{2},6,9}{z}}{525000 \pi ^4}
                    +\frac{16 \alpha _s^2 m_Q^{10} z^6 \, \pFq{4}{3}{1,1,3,5}{\frac{7}{2},6,10}{z}}{759375 \pi ^4}
                    +\frac{16 \alpha _s^2 m_Q^{10} z^6 \, \pFq{4}{3}{1,1,3,5}{\frac{7}{2},6,11}{z}}{590625 \pi ^4}
                    -\frac{17 \alpha _s^2 m_Q^{10} z^6 \, \pFq{4}{3}{1,1,3,5}{\frac{7}{2},7,9}{z}}{1575000 \pi ^4}
                    +\frac{47 \alpha _s^2 m_Q^{10} z^6 \, \pFq{4}{3}{1,1,3,5}{\frac{7}{2},7,10}{z}}{15946875 \pi ^4}
                    +\frac{2 \alpha _s^2 m_Q^{10} z^6 \, \pFq{4}{3}{1,1,3,5}{4,\frac{9}{2},11}{z}}{12403125 \pi ^4}
                    +\frac{\alpha _s^2 m_Q^{10} z^6 \, \pFq{4}{3}{1,1,3,6}{\frac{7}{2},5,11}{z}}{50625 \pi ^4}
                    +\frac{\alpha _s^2 m_Q^{10} z^6 \, \pFq{4}{3}{1,1,3,6}{\frac{7}{2},7,10}{z}}{25515000 \pi ^4}
                    -\frac{\alpha _s^2 m_Q^{10} z^6 \, \pFq{4}{3}{1,1,3,6}{\frac{7}{2},7,11}{z}}{759375 \pi ^4}
                    -\frac{\alpha _s^2 m_Q^{10} z^6 \, \pFq{4}{3}{1,1,4,7}{\frac{9}{2},6,11}{z}}{1575000 \pi ^4}\,,
                \end{autobreak}\\
                \Pi^{\GGb}_{J_{\mu\nu\rho}^3}=&\frac{\alpha _s \GGb m_Q^6 (z-1) (4 z (24 z (10 z+21)-121)-35) \, \pFq{3}{2}{1,1,1}{\frac{3}{2},3}{z}}{46080 \pi ^2 z}\\
                &+\frac{\alpha _s \GGb m_Q^6 (8 z (z (48 z (5 z-32)-301)+57)+35) \, \pFq{3}{2}{1,1,2}{\frac{5}{2},4}{z}}{414720 \pi ^2 }\,,\\
                \Pi^{\GGGb}_{J_{\mu\nu\rho}^3}=&-\frac{\alpha _s \GGGb m_Q^4 (z-1) (4 z (5028 z-2797)-1395) \, \pFq{3}{2}{1,1,1}{\frac{3}{2},3}{z}}{2211840 \pi ^3 z}\\
                &-\frac{\alpha _s \GGGb m_Q^4 (8 z (z (2208 z-8347)+1259)+1395) \, \pFq{3}{2}{1,1,2}{\frac{5}{2},4}{z}}{19906560 \pi ^3}\\
                &-\frac{5 \alpha _s \GGGb m_Q^4 z (z-1) \left(3 \log \left(\frac{m_Q^2}{\mu^2}\right)-2\right) \, \pFq{2}{1}{1,1}{\frac{2}{2}}{z}}{2592 \pi ^3}\,,\\
                \Pi^{R,\,\text{fig.2(g)}}_{J_{\mu\nu\rho}^3}=&\frac{\alpha _s \GGGb m_Q^4\left(9 \left(-6 z^2+z+5\right)
                \, \pFq{3}{2}{1,1,1}{\frac{3}{2},3}{z}+z \left(48 z^2+2 z+5\right)
                \, \pFq{3}{2}{1,1,2}{\frac{5}{2},4}{z}\right)}{31104 \pi ^3}\\
                &-\frac{5 \alpha _s \GGGb m_Q^4 z (z-1) \left(3 \log \left(\frac{m_Q^2}{\mu^2}\right)-2\right) \, \pFq{2}{1}{1,1}{\frac{2}{2}}{z}}{2592 \pi ^3}\,,\\
                \Pi^{\GGb^2}_{J_{\mu\nu\rho}^3}=&\frac{\GGb^2}{432} m_Q^2 z (2 z-5) \, \pFq{2}{1}{1,1}{\frac{2}{2}}{z}\,,\\
                \begin{autobreak}
                    \rho^{pert}_{J_{\mu\nu\rho}^3}=
                    \frac{21 \alpha _s^2 m_Q^{10} z^4 \MeijerG{2,0}{2,2}{\frac{5}{2},9}{1,1}{\frac{1}{z}}}{10 \pi ^{7/2}}
                    -\frac{21 \alpha _s^2 m_Q^{10} z^4 \MeijerG{2,0}{2,2}{\frac{5}{2},9}{1,2}{\frac{1}{z}}}{50 \pi ^{7/2}}
                    +\frac{448 \alpha _s^2 m_Q^{10} z^6 \MeijerG{2,0}{2,2}{\frac{5}{2},10}{1,2}{\frac{1}{z}}}{5 \pi ^{7/2}}
                    -\frac{2688 \alpha _s^2 m_Q^{10} z^6 \MeijerG{2,0}{2,2}{\frac{5}{2},11}{1,2}{\frac{1}{z}}}{5 \pi ^{7/2}}
                    +\frac{546 \alpha _s^2 m_Q^{10} z^5 \MeijerG{2,0}{2,2}{\frac{7}{2},10}{1,2}{\frac{1}{z}}}{25 \pi ^{7/2}}
                    -\frac{557 \alpha _s^2 m_Q^{10} z^6 \MeijerG{2,0}{2,2}{\frac{7}{2},10}{1,3}{\frac{1}{z}}}{25 \pi ^{7/2}}
                    -\frac{147 \alpha _s^2 m_Q^{10} z^5 \MeijerG{2,0}{2,2}{\frac{7}{2},10}{1,3}{\frac{1}{z}}}{25 \pi ^{7/2}}
                    +\frac{6028 \alpha _s^2 m_Q^{10} z^6 \MeijerG{2,0}{2,2}{\frac{7}{2},11}{1,3}{\frac{1}{z}}}{25 \pi ^{7/2}}
                    +\frac{924 \alpha _s^2 m_Q^{10} z^6 \MeijerG{2,0}{2,2}{\frac{9}{2},11}{1,3}{\frac{1}{z}}}{25 \pi ^{7/2}}
                    -\frac{153 \alpha _s^2 m_Q^{10} z^6 \MeijerG{2,0}{2,2}{\frac{9}{2},11}{1,4}{\frac{1}{z}}}{5 \pi ^{7/2}}
                    +\frac{21 \alpha _s^2 m_Q^{10} z^6 \MeijerG{2,0}{2,2}{\frac{9}{2},11}{1,5}{\frac{1}{z}}}{50 \pi ^{7/2}}
                    +\frac{21 \alpha _s^2 m_Q^{10} z^6 \MeijerG{2,0}{2,2}{\frac{9}{2},11}{1,6}{\frac{1}{z}}}{50 \pi ^{7/2}}
                    -\frac{96 \alpha _s^2 m_Q^{10} z^5 \MeijerG{3,0}{3,3}{\frac{5}{2},4,9}{1,2,3}{\frac{1}{z}}}{5 \pi ^{7/2}}
                    -\frac{86 \alpha _s^2 m_Q^{10} z^5 \MeijerG{3,0}{3,3}{\frac{5}{2},5,8}{1,1,4}{\frac{1}{z}}}{25 \pi ^{7/2}}
                    -\frac{9 \alpha _s^2 m_Q^{10} z^4 \MeijerG{3,0}{3,3}{\frac{5}{2},5,8}{1,1,4}{\frac{1}{z}}}{25 \pi ^{7/2}}
                    -\frac{128 \alpha _s^2 m_Q^{10} z^5 \MeijerG{3,0}{3,3}{\frac{5}{2},5,8}{1,2,3}{\frac{1}{z}}}{25 \pi ^{7/2}}
                    -\frac{1904 \alpha _s^2 m_Q^{10} z^5 \MeijerG{3,0}{3,3}{\frac{5}{2},5,9}{1,1,4}{\frac{1}{z}}}{25 \pi ^{7/2}}
                    +\frac{1664 \alpha _s^2 m_Q^{10} z^6 \MeijerG{3,0}{3,3}{\frac{5}{2},5,9}{1,2,4}{\frac{1}{z}}}{25 \pi ^{7/2}}
                    +\frac{288 \alpha _s^2 m_Q^{10} z^5 \MeijerG{3,0}{3,3}{\frac{5}{2},5,9}{1,2,4}{\frac{1}{z}}}{25 \pi ^{7/2}}
                    +\frac{192 \alpha _s^2 m_Q^{10} z^5 \MeijerG{3,0}{3,3}{\frac{5}{2},5,10}{1,1,4}{\frac{1}{z}}}{\pi ^{7/2}}
                    -\frac{384 \alpha _s^2 m_Q^{10} z^6 \MeijerG{3,0}{3,3}{\frac{5}{2},5,10}{1,2,4}{\frac{1}{z}}}{5 \pi ^{7/2}}
                    -\frac{1536 \alpha _s^2 m_Q^{10} z^6 \MeijerG{3,0}{3,3}{\frac{5}{2},5,11}{1,2,4}{\frac{1}{z}}}{\pi ^{7/2}}
                    +\frac{322 \alpha _s^2 m_Q^{10} z^5 \MeijerG{3,0}{3,3}{\frac{5}{2},6,7}{1,2,3}{\frac{1}{z}}}{25 \pi ^{7/2}}
                    -\frac{36 \alpha _s^2 m_Q^{10} z^6 \MeijerG{3,0}{3,3}{\frac{5}{2},6,8}{1,2,4}{\frac{1}{z}}}{25 \pi ^{7/2}}
                    +\frac{16 \alpha _s^2 m_Q^{10} z^5 \MeijerG{3,0}{3,3}{\frac{5}{2},6,8}{1,2,4}{\frac{1}{z}}}{25 \pi ^{7/2}}
                    +\frac{2864 \alpha _s^2 m_Q^{10} z^6 \MeijerG{3,0}{3,3}{\frac{5}{2},6,9}{1,2,4}{\frac{1}{z}}}{25 \pi ^{7/2}}
                    -\frac{544 \alpha _s^2 m_Q^{10} z^6 \MeijerG{3,0}{3,3}{\frac{5}{2},6,10}{1,2,4}{\frac{1}{z}}}{\pi ^{7/2}}
                    -\frac{1504 \alpha _s^2 m_Q^{10} z^6 \MeijerG{3,0}{3,3}{3,\frac{7}{2},10}{1,2,4}{\frac{1}{z}}}{25 \pi ^{7/2}}
                    +\frac{32 \alpha _s^2 m_Q^{10} z^6 \MeijerG{3,0}{3,3}{3,\frac{7}{2},11}{1,2,4}{\frac{1}{z}}}{5 \pi ^{7/2}}
                    -\frac{224 \alpha _s^2 m_Q^{10} z^6 \MeijerG{3,0}{3,3}{3,\frac{7}{2},11}{1,2,5}{\frac{1}{z}}}{5 \pi ^{7/2}}
                    +\frac{73 \alpha _s^2 m_Q^{10} z^6 \MeijerG{3,0}{3,3}{\frac{7}{2},4,10}{1,2,5}{\frac{1}{z}}}{25 \pi ^{7/2}}
                    +\frac{396 \alpha _s^2 m_Q^{10} z^6 \MeijerG{3,0}{3,3}{\frac{7}{2},4,11}{1,2,5}{\frac{1}{z}}}{25 \pi ^{7/2}}
                    +\frac{336 \alpha _s^2 m_Q^{10} z^6 \MeijerG{3,0}{3,3}{\frac{7}{2},4,11}{1,2,6}{\frac{1}{z}}}{25 \pi ^{7/2}}
                    +\frac{336 \alpha _s^2 m_Q^{10} z^6 \MeijerG{3,0}{3,3}{\frac{7}{2},5,11}{1,3,6}{\frac{1}{z}}}{25 \pi ^{7/2}}
                    -\frac{172 \alpha _s^2 m_Q^{10} z^6 \MeijerG{3,0}{3,3}{\frac{7}{2},6,9}{1,2,5}{\frac{1}{z}}}{25 \pi ^{7/2}}
                    -\frac{18 \alpha _s^2 m_Q^{10} z^5 \MeijerG{3,0}{3,3}{\frac{7}{2},6,9}{1,2,5}{\frac{1}{z}}}{25 \pi ^{7/2}}
                    -\frac{104 \alpha _s^2 m_Q^{10} z^6 \MeijerG{3,0}{3,3}{\frac{7}{2},6,9}{1,3,5}{\frac{1}{z}}}{25 \pi ^{7/2}}
                    -\frac{9 \alpha _s^2 m_Q^{10} z^5 \MeijerG{3,0}{3,3}{\frac{7}{2},6,9}{1,3,5}{\frac{1}{z}}}{25 \pi ^{7/2}}
                    -\frac{3808 \alpha _s^2 m_Q^{10} z^6 \MeijerG{3,0}{3,3}{\frac{7}{2},6,10}{1,2,5}{\frac{1}{z}}}{25 \pi ^{7/2}}
                    +\frac{896 \alpha _s^2 m_Q^{10} z^6 \MeijerG{3,0}{3,3}{\frac{7}{2},6,10}{1,3,5}{\frac{1}{z}}}{25 \pi ^{7/2}}
                    +\frac{384 \alpha _s^2 m_Q^{10} z^6 \MeijerG{3,0}{3,3}{\frac{7}{2},6,11}{1,2,5}{\frac{1}{z}}}{\pi ^{7/2}}
                    +\frac{2304 \alpha _s^2 m_Q^{10} z^6 \MeijerG{3,0}{3,3}{\frac{7}{2},6,11}{1,3,5}{\frac{1}{z}}}{5 \pi ^{7/2}}
                    -\frac{306 \alpha _s^2 m_Q^{10} z^6 \MeijerG{3,0}{3,3}{\frac{7}{2},7,9}{1,3,5}{\frac{1}{z}}}{25 \pi ^{7/2}}
                    +\frac{752 \alpha _s^2 m_Q^{10} z^6 \MeijerG{3,0}{3,3}{\frac{7}{2},7,10}{1,3,5}{\frac{1}{z}}}{25 \pi ^{7/2}}
                    +\frac{2 \alpha _s^2 m_Q^{10} z^6 \MeijerG{3,0}{3,3}{\frac{7}{2},7,10}{1,3,6}{\frac{1}{z}}}{25 \pi ^{7/2}}
                    -\frac{672 \alpha _s^2 m_Q^{10} z^6 \MeijerG{3,0}{3,3}{\frac{7}{2},7,11}{1,3,6}{\frac{1}{z}}}{25 \pi ^{7/2}}
                    -\frac{21 \alpha _s^2 m_Q^{10} z^6 \MeijerG{3,0}{3,3}{4,\frac{9}{2},11}{1,2,7}{\frac{1}{z}}}{50 \pi ^{7/2}}
                    +\frac{12 \alpha _s^2 m_Q^{10} z^6 \MeijerG{3,0}{3,3}{4,\frac{9}{2},11}{1,3,5}{\frac{1}{z}}}{25 \pi ^{7/2}}
                    -\frac{21 \alpha _s^2 m_Q^{10} z^6 \MeijerG{3,0}{3,3}{\frac{9}{2},6,11}{1,4,7}{\frac{1}{z}}}{50 \pi ^{7/2}}\,,
                \end{autobreak}\\
                \rho^{\GGb}_{J_{\mu\nu\rho}^3}=&\frac{\alpha _s \GGb m_Q^6 (4 z (24 z (10 z+11)-625)+449) \MeijerG{2,0}{2,2}{\frac{3}{2},3}{1,1}{\frac{1}{z}}}{46080 \pi ^{3/2}}\\
                &+\frac{\alpha _s \GGb m_Q^6 \left((8 z (z (48 z (5 z-32)-301)+57)+35) \MeijerG{2,0}{2,2}{\frac{5}{2},4}{1,2}{\frac{1}{z}}+70 \MeijerG{2,0}{2,2}{\frac{5}{2},4}{2,2}{\frac{1}{z}}\right)}{92160 \pi ^{3/2}}\,,\\
                \rho^{\GGGb}_{J_{\mu\nu\rho}^3}=&-\frac{5\alpha _s \GGGb m_Q^4 z (z-1) \left(3 \log \left(\frac{m_Q^2}{\mu^2}\right)-2\right) \MeijerG{1,0}{1,1}{\frac{5}{2}}{1}{\frac{1}{z}}}{3456 \pi ^{5/2}}\\
                &-\frac{\alpha _s \GGGb m_Q^4 \big(8 z (z (2208 z-8347)+1259)+1395\big) \MeijerG{2,0}{2,2}{\frac{5}{2},4}{1,2}{\frac{1}{z}}}{4423680 \pi ^{5/2}}\\
                &-\frac{\alpha _s \GGGb m_Q^4 (z-1) (4 z (5028 z-2797)-1395) \MeijerG{2,0}{2,2}{\frac{3}{2},3}{1,1}{\frac{1}{z}}}{2211840 \pi ^{5/2} z}\,,\\
                \rho^{R,\,\text{fig.2(g)}}_{J_{\mu\nu\rho}^3}=& \frac{\alpha _s \GGGb m_Q^4\left(2 \left(-6 z^2+z+5\right)
                 \MeijerG{2,0}{2,2}{\frac{3}{2},3}{1,1}{\frac{1}{z}}+z \left(48 z^2+2 z+5\right)
                 \MeijerG{2,0}{2,2}{\frac{5}{2},4}{1,2}{\frac{1}{z}}\right)}{6912 \pi ^{5/2}}\\
                &-\frac{5\alpha _s \GGGb m_Q^4 z (z-1) \left(3 \log \left(\frac{m_Q^2}{\mu^2}\right)-2\right) \MeijerG{1,0}{1,1}{\frac{5}{2}}{1}{\frac{1}{z}}}{3456 \pi ^{5/2}}\,,\\
                \rho^{\GGb^2}_{J_{\mu\nu\rho}^3}=&\frac{\GGb^2}{576} m_Q^2 z (2 z-5) \pi^{\frac{1}{2}} \MeijerG{1,0}{1,1}{\frac{5}{2}}{1}{\frac{1}{z}}\,,
            \end{align*}
        \item[$\bullet$] $J_{\mu\nu\rho}^4$:
            \begin{align*}
                \Pi^{pert}_{J_{\mu\nu\rho}^4}=& \Pi^{pert}_{J_{\mu\nu\rho}^3},\quad \Pi^{\GGb}_{J_{\mu\nu\rho}^4}=-\Pi^{\GGb}_{J_{\mu\nu\rho}^3},\quad \Pi^{\GGb^4}_{J_{\mu\nu\rho}^2}=\Pi^{\GGb^2}_{J_{\mu\nu\rho}^3}\,,\\
                \Pi^{\GGGb}_{J_{\mu\nu\rho}^4}=&\frac{\alpha _s \GGGb m_Q^4 (z-1) (4 z (6468 z-2617)-1395) \, \pFq{3}{2}{1,1,1}{\frac{3}{2},3}{z}}{2211840 \pi ^3 z}\\
                &+\frac{\alpha _s \GGGb m_Q^4 (8 z (z (5808 z-15907)+1169)+1395) \, \pFq{3}{2}{1,1,2}{\frac{5}{2},4}{z}}{19906560 \pi ^3}\\
                &+\frac{\alpha _s \GGGb m_Q^4 z (z-1) \left(3 \log \left(\frac{m_Q^2}{\mu^2}\right)-5\right) \, \pFq{2}{1}{1,1}{\frac{5}{2}}{z}}{1296 \pi ^3}\,,\\
                \Pi^{R,\,\text{fig.2(g)}}_{J_{\mu\nu\rho}^4}=& -\frac{\alpha _s \GGGb m_Q^4\left(9 \left(12 z^2-47 z+35\right)
                \, \pFq{3}{2}{1,1,1}{\frac{3}{2},3}{z}+5 z \left(24 z^2-8 z+7\right)
                \, \pFq{3}{2}{1,1,2}{\frac{5}{2},4}{z}\right)}{124416 \pi ^3}\\
                &+\frac{\alpha _s \GGGb m_Q^4 z (z-1) \left(3 \log \left(\frac{m_Q^2}{\mu^2}\right)-5\right) \, \pFq{2}{1}{1,1}{\frac{5}{2}}{z}}{1296 \pi ^3}\,,\\
                \rho^{pert}_{J_{\mu\nu\rho}^4}=&\rho^{pert}_{J_{\mu\nu\rho}^3},\quad \rho^{\GGb}_{J_{\mu\nu\rho}^4}=-\rho^{\GGb}_{J_{\mu\nu\rho}^3},\quad \rho^{\GGb^2}_{J_{\mu\nu\rho}^4}=\rho^{\GGb^2}_{J_{\mu\nu\rho}^3}\,,\\
                \rho^{\GGGb}_{J_{\mu\nu\rho}^4}=&\frac{\alpha _s \GGGb m_Q^4 z (z-1) \left(3 \log \left(\frac{m_Q^2}{\mu^2}\right)-5\right) \MeijerG{1,0}{1,1}{\frac{5}{2}}{1}{\frac{1}{z}}}{1728\pi ^{5/2}}\\
                &+\frac{\alpha _s \GGGb m_Q^4 (8 z (z (5808 z-15907)+1169)+1395) \MeijerG{2,0}{2,2}{\frac{5}{2},4}{1,2}{\frac{1}{z}}}{4423680 \pi ^{5/2}}\\
                &+\frac{\alpha _s \GGGb m_Q^4 (z-1) (4 z (6468 z-2617)-1395) \MeijerG{2,0}{2,2}{\frac{3}{2},3}{1,1}{\frac{1}{z}}}{2211840 \pi ^{5/2} z}\,,\\
                \rho^{R,\,\text{fig.2(g)}}_{J_{\mu\nu\rho}^4}=&-\frac{\alpha _s \GGGb m_Q^4\left(2 \left(12 z^2-47 z+35\right)
                \MeijerG{2,0}{2,2}{\frac{3}{2},3}{1,1}{\frac{1}{z}}+5 z \left(24 z^2-8 z+7\right)
                \MeijerG{2,0}{2,2}{\frac{5}{2},4}{1,2}{\frac{1}{z}}\right)}{27648 \pi ^{5/2}}\\
                &+\frac{\alpha _s \GGGb m_Q^4 z (z-1) \left(3 \log \left(\frac{m_Q^2}{\mu^2}\right)-5\right) \MeijerG{1,0}{1,1}{\frac{5}{2}}{1}{\frac{1}{z}}}{1728\pi ^{5/2}}\,.
            \end{align*}
        \end{itemize}
\end{itemize}

% \newpage
% \nocite{*}
% \bibliography{2g2q_ref}

\end{document}